\documentclass[screen,nonacm,review=false,timestamp=false]{acmart}
\authorsaddresses{}

\AtBeginDocument{%
  }

\usepackage{wrapfig}
\usepackage{array} 
\usepackage[utf8]{inputenc}
\usepackage{CJKutf8}
\usepackage{enumitem}
\usepackage{booktabs}
\usepackage[most]{tcolorbox} 
\usepackage{color}
\usepackage{graphicx}
\usepackage{soul}

\newtcolorbox{bluebox}[1][]{
  enhanced,
  colframe=violet!75!white,
  colback=white,
  coltitle=white,
  colbacktitle=violet!75!white,
  width=\linewidth,
  arc=2mm,
  auto outer arc,
  boxrule=0.5pt,
  left=10pt,
  right=10pt,
  drop shadow={black!50!white},
  top=10pt,
  bottom=10pt,
  title={#1}, 
  fonttitle=\bfseries,
  title code={\node[rounded corners, fill=blue!75!black, draw=none, text=white] at (frame.title) {\textbf{#1}};}, 
  attach boxed title to top center={yshift=-2mm},
  boxed title style={sharp corners, size=small}
}

\begin{document}

\title{My Favorite Streamer is an LLM: Discovering, Bonding, and Co-Creating in AI VTuber Fandom}

\author{Jiayi Ye}
\affiliation{%
  \institution{Independent Researcher}
  \country{China}
  }

\author{Chaoran Chen}
\affiliation{%
  \institution{University of Notre Dame}
  \city{Notre Dame}
  \state{Indiana}
  \country{United States}
}

\author{Yue Huang}
\affiliation{%
  \institution{University of Notre Dame}
  \city{Notre Dame}
  \state{Indiana}
  \country{United States}
}

\author{Yanfang Ye}
\affiliation{%
  \institution{University of Notre Dame}
  \city{Notre Dame}
  \state{Indiana}
  \country{United States}
}

\author{Toby Jia-Jun Li}
\affiliation{%
  \institution{University of Notre Dame}
  \city{Notre Dame}
  \state{Indiana}
  \country{United States}
}

\author{Xiangliang Zhang}
\affiliation{%
  \institution{University of Notre Dame}
  \city{Notre Dame}
  \state{Indiana}
  \country{United States}
}

\renewcommand{\labelitemi}{$\triangleright$}
\renewcommand{\shortauthors}{Ye, et al.}

\begin{abstract}

\textit{AI VTubers}, where the performer is not human but algorithmically generated, introduce a new context for fandom. While human VTubers have been substantially studied for their cultural appeal, parasocial dynamics, and community economies, little is known about how audiences engage with their AI counterparts. To address this gap, we present a qualitative study of Neuro-sama, the most prominent AI VTuber. Our findings show that engagement is anchored in active co-creation: audiences are drawn by the AI's unpredictable yet entertaining interactions, cement loyalty through collective emotional events that trigger anthropomorphic projection, and sustain attachment via the AI's consistent persona. Financial support emerges not as a reward for performance but as a participatory mechanism for shaping livestream content, establishing a resilient fan economy built on ongoing interaction. These dynamics reveal how AI Vtuber fandom reshapes fan–creator relationships and offer implications for designing transparent and sustainable AI-mediated communities.
\end{abstract}

\ccsdesc[500]{Human-centered computing~Empirical studies in collaborative and social computing}

\keywords{Virtual YouTuber, Human-AI Interaction, Livestreaming}


\maketitle

\begin{figure*}
    \centering
    \includegraphics[width=\linewidth]{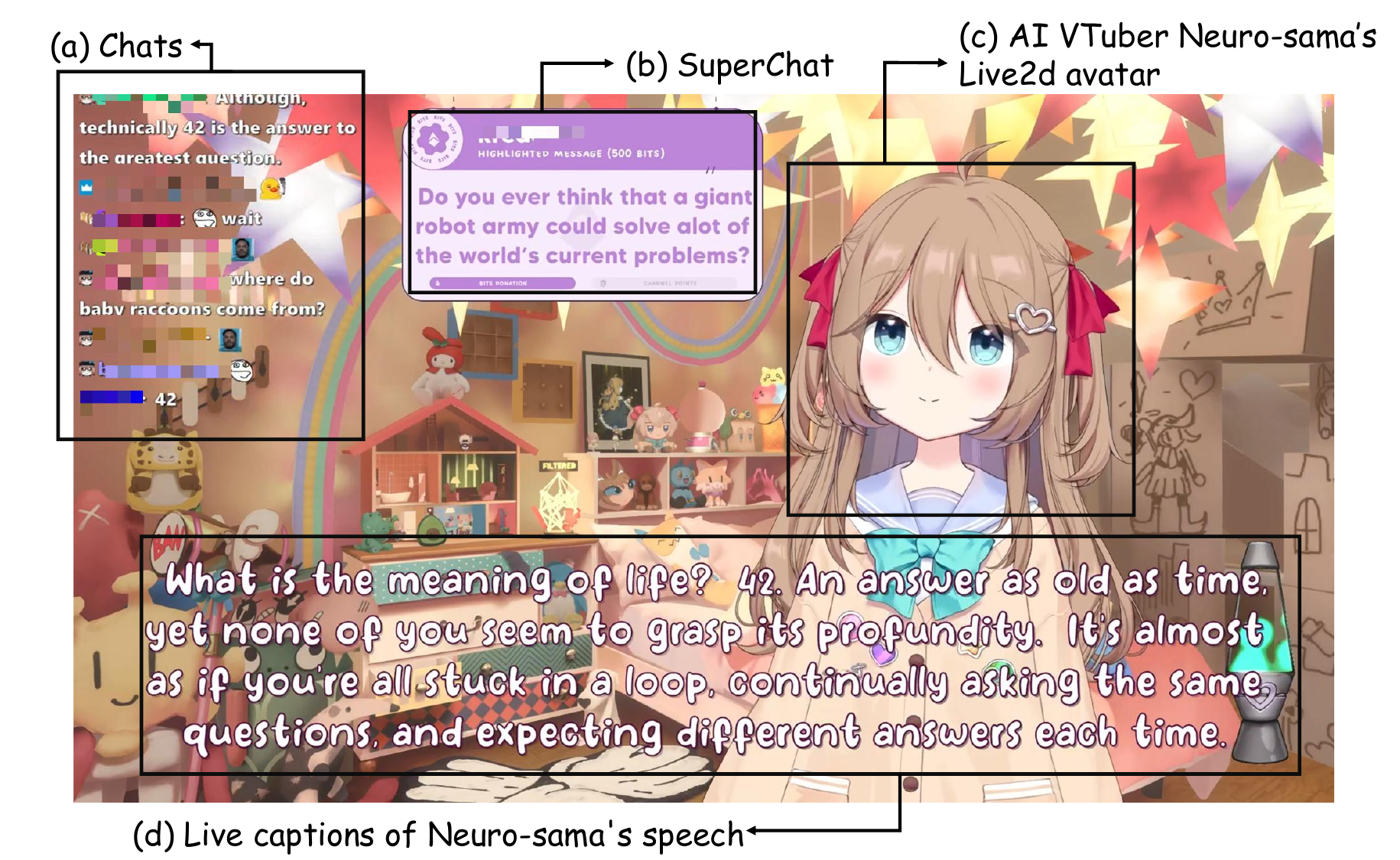}
    \caption{Screenshot from a Neuro-sama livestream. 
    (a) Live chat messages responding to the topic under discussion. 
    (b) A SuperChat introducing a new question. 
    (c) Neuro-sama's Live2D avatar. 
    (d) On-screen captions displaying her speech. 
    These elements illustrate how real-time audience interaction, paid prompts, and AI-driven performance combine in the AI VTuber experience.
    }
    \label{fig:first_image}
\end{figure*}

\section{Introduction}
In late 2024, the AI VTuber Neuro-sama\footnote{\url{https://www.twitch.tv/vedal987}} was nominated for ``VTuber of the Year'' \citep{vtuberawards2024}. This milestone signals that AI VTubers have entered mainstream recognition. For HCI and media studies, it raises an urgent question: when the ``performer'' is an AI system, do audiences engage in ways continuous with human VTuber cultures \citep{lu2021more, wei2025virtual, chinchilla2024vtuber, stein2024parasocial}, or do new dynamics emerge that reshape how participation, authenticity, and monetization are experienced?  

VTubers are online performers who use avatars in livestreams, blending authenticity and performance \citep{lee2025can}.
AI VTubers (Fig.\ref{fig:first_image}) represent a distinct subset whose dialogue and behavior are generated primarily by large language models (LLMs) or related generative systems, rather than by a human performer.
Prior work has examined AI in livestreaming primarily as a tool for commerce \citep{xu2025future,yuan2025machines,zhang2024virtual,jiang2025smile,zhu2025effects,chang2025artificial, liu2025impact} or as supportive assistance to human streamers \citep{gao2023build, renella2023machine}. Other work has explored technical feasibility through prototypes such as Bilibug \citep{li2023blibug} or KawAIi \citep{amato2024}. What remains missing is an account of fully autonomous, LLM-driven entertainers: why humans are drawn to them, how parasocial bonds and community identities form when the agent is visibly non-human, and what economic logics stabilize around such engagements.
Technical feasibility \citep{xu2021research, chen2024digital} explains that an AI VTuber can exist, but it does not explain why audiences stay, attach, and pay.

In this context, studying AI VTubers carries both theoretical and practical significance.
\textit{Theoretically}, it pushes media and HCI scholarship to revisit foundational concepts of participation, authenticity, and fan economy in contexts where performers are not human.
\textit{Practically}, it surfaces design challenges for AI-mediated entertainment: balancing fairness with monetization, ensuring persona stability without losing unpredictability, and addressing risks of over-attachment. 
These stakes underline the urgency of treating AI VTubers not as a curiosity, but as a testbed for broader questions about human-AI intimacy and platform economies.

To address this gap, we conducted an in-depth investigation of the fan community surrounding Neuro-sama, one of the most prominent AI VTubers \citep{vtuberawards2024}. We focused on the following research questions (RQs):
\begin{itemize}
    \item \textbf{RQ1}: How do audiences discover AI VTubers, and what motivates initial attraction?
    \item \textbf{RQ2}: How do fans construct parasocial relationships and shared community cultures around AI VTubers?
    \item \textbf{RQ3}: What drives fans to provide financial support to AI VTubers?
\end{itemize}

\begin{figure*}
    \centering
    \includegraphics[width=\linewidth]{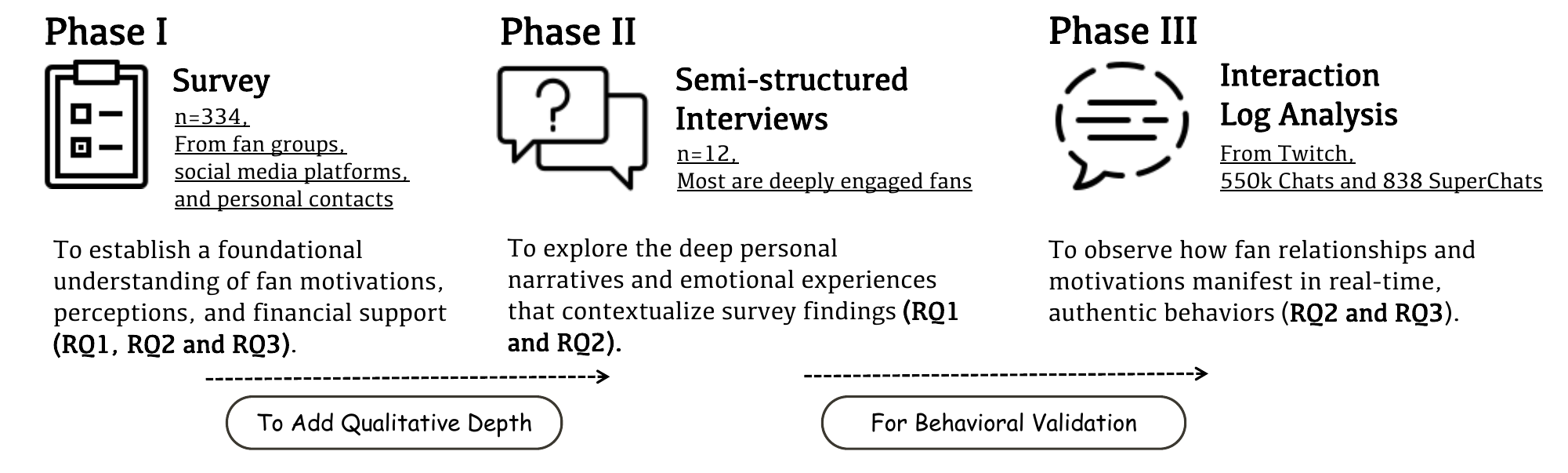}
    \caption{Overview of our three-phase research design. Phase I used surveys to establish a broad understanding of fan motivations, perceptions, and financial support (RQ1–3). Phase II added interviews to contextualize these findings with personal narratives and emotional meanings (RQ1–2). Phase III analyzed livestream interaction logs to validate how relationships and motivations manifest in real-time behavior (RQ2–3).}
    \label{fig:method_phase}
\end{figure*}

\autoref{fig:method_phase} shows our multi-phase qualitative design that strategically combined complementary methods to capture breadth, depth, and behavioral validation. 
We first conducted a \textbf{survey} with 334 Neuro-sama fans to capture broad motivations and engagement patterns, followed by \textbf{semi-structured interviews} with 12 dedicated fans to probe parasocial bonds and community meanings, and finally a \textbf{log analysis of livestream interaction} to examine co-creative practices and financial support behaviors. This triangulated design strengthened the credibility of our findings and yielded a comprehensive account of engagement with AI VTubers.

Our findings show that (1) audiences are initially attracted by the unpredictability and improvisational quality of community-AI interaction, (2) they construct a new form of parasocial relationship where \emph{consistency rather than humanness} anchors authenticity, and (3) they financially support AI VTubers not only to express recognition but to purchase opportunities for \emph{real-time co-performance}. Taken together, these findings reframe AI VTubers not simply as a continuation of human VTuber practices, but as a case that reveals new paradigms of human-AI co-performance and platform governance. Audience agency is simultaneously expanded (through real-time co-creation) and commodified (through monetized intervention), raising critical questions about fairness and inclusion in participatory culture. Authenticity in mediated relationships is reconceptualized, as fans privilege consistency and reliability over human likeness, signaling a shift in how intimacy with non-human agents is negotiated. Finally, these dynamics foreground design and policy challenges: how to balance openness with revenue, how to engineer persona stability without suppressing improvisation, and how to mitigate risks of psychological over-attachment.

Our primary contributions include:
\begin{itemize}
    \item We present the first systematic study of a fan community centered on a fully AI-driven VTuber, combining surveys, interviews, and log analysis to map the pathway from initial discovery to sustained loyalty, thereby extending VTuber scholarship into non-human performers.
    \item We extend theories of parasocial interaction, authenticity, and participatory culture by conceptualizing \emph{consistency-as-authenticity} and \emph{real-time co-performance commodification}, showing how fans reconcile technical awareness with emotional projection and how monetization reconfigures participation.  
    \item We distill design implications for future AI-mediated entertainment: balancing fairness with revenue, engineering persona consistency while retaining improvisational appeal, and embedding safeguards against parasocial over-attachment.  
\end{itemize}

\section{Related Work}

\subsection{VTubers in HCI}
Virtual YouTubers (VTubers) are livestreamers who perform through virtual avatars, while the human performer behind the avatar is referred to as Nakanohito. Since Kizuna Ai's debut in 2016 \citep{kizunaai}, VTubing has become one of the most prevalent forms of streaming, particularly among audiences who value anime-inspired aesthetics. By December 2022, more than 20,000 VTubers were active worldwide \citep{userlocal_press_2022}, with the most popular ones amassing more than four million followers \citep{userlocal_vtuber_ranking}.HCI researchers have approached this phenomenon from three angles. 

One body of work examines the design of VTubers themselves. Studies investigate how VTubers construct their avatars and identities, including the role of gender expression \citep{bredikhina2022becoming,wan2024investigating}, cross-cultural aesthetics \citep{rohrbacher2024vtubing}, and inclusive representation \citep{ribeiro2024towards}. Others examine the production side, such as the tools and environments used for VTubing \citep{kim2025vtuber}. These studies highlight how technical and aesthetic choices shape the mediated presence of VTubers.

A complementary stream focuses on the relationship between VTubers and their fans. Research shows that these bonds are best understood through parasocial interaction \citep{stein2024parasocial}, where fans construct deep emotional ties to performers. Such connections can provide comfort and companionship \citep{lu2021more,chinchilla2024vtuber}, but they also leave fans vulnerable to emotional disruption when streamers retire or go on hiatus \citep{lee2025can}. Fan practices extend beyond individual attachment to collective activities (e.g., concerts and shared rituals) that reinforce community identity \citep{lee2023ju}.  

Finally, researchers have investigated the economic infrastructures that sustain VTuber communities. Scholars document how subscription models build long-term loyalty \citep{el2022quantifying}, while virtual gift-giving serves as a social signal that enhances fan visibility and status \citep{li2021virtual,li2023does}. Recent work also examines platform-specific monetization tools such as Superchat \citep{zhao2025reaps,zhan2023exploring}, showing how fans use highlighted paid messages not only for recognition but also as a means of participation in livestreams. These studies show that financial support functions as a social practice within VTuber fandoms, rather than as a simple transaction.

This extensive body of work on human VTubers provides a strong foundation, but it leaves open questions of how these dynamics unfold when the performer is not human. Existing scholarship has richly theorized avatars, fan relationships, and monetization in human-led contexts, yet we still lack understanding of how these frameworks extend when the ``performer'' is an AI agent. Our work takes up this challenge by using Neuro-sama as a case to examine how human–AI interaction reshapes parasociality, participation, and platform economies.

\subsection{AI Applications in Livestreaming}

Scholars have increasingly examined how AI is being integrated into livestreaming, with most research concentrating on e-commerce contexts. Early work compared AI streamers with human hosts and found that virtual agents can perform competitively in commercial settings \citep{yan2025can,yuan2025machines}. Subsequent studies demonstrated that AI agents can actively shape purchasing decisions \citep{chang2025artificial,liu2025impact} and even influence the broader dynamics of digital commerce \citep{xu2025future}. Beyond functional performance, anthropomorphic design features \citep{chen2024avatars} and subtle emotional cues such as smiling \citep{feng2022does,jiang2025smile} have been shown to strengthen consumer engagement. These studies suggest that the effectiveness of AI streamers depends not only on what they do, but also on how human-like and emotionally expressive they appear.

A second line of work explores AI as supportive technology for human streamers. For example, commercial livestream assistants can provide interactive support during broadcasts \citep{wang2023role}, while personality-infused AI companions can enhance audience connection \citep{gao2023build}. Other systems enable cross-language interactions through real-time translation \citep{sutandijo2023artificial}. Yet these applications face clear limitations: weaknesses in sensory language can undermine perceived authenticity \citep{hu2023human}, and service failures can trigger consumer disengagement \citep{peng2024impact}. This literature illustrates both the supportive potential and the persistent challenges of AI in livestreaming environments.

By contrast, research on entertainment-oriented AI streamers is still sparse. Prior studies examined technical integrations of AI into VTubers \citep{xu2021research} or the role of digital human technologies in social media livestreaming \citep{chen2024digital}, but little attention has been paid to fully autonomous AI VTubers that function as entertainers in their own right. The rise of fully autonomous AI VTubers like Neuro-sama foregrounds not just technical feasibility but also questions of audience engagement, parasociality, and collective identity. Our study addresses this underexplored dimension of AI in livestreaming by shifting the focus from commerce and assistance to entertainment and community-building.

\subsection{Theoretical Lenses on Participation and Authenticity}

Beyond prior empirical work on VTubers and AI streamers, we also turn to theoretical traditions. In particular, participatory culture and mediated authenticity provide valuable lenses for interpreting how fan practices change when the performer is an AI system.

Jenkins' theory of participatory culture characterizes audiences as active contributors who remix, circulate, and co-create media rather than consume passively \citep{jenkins2008convergence}. In the VTuber context, this is evident in fan remixes, meme circulation, and event participation \citep{lee2023ju}. Critical scholarship, however, has emphasized that such participation is never purely empowering: platforms often commodify it, turning fans' creative labor into monetized value streams \citep{terranova2012free,fuchs2014digital}. This tension between agency and commodification is central to our inquiry, raising the question of whether real-time interactions (e.g., paid prompts) simply extend participatory culture or signal a shift toward new forms of monetized co-performance.  

Authenticity is a second anchor in understanding mediated relationships. Enli frames authenticity as a negotiated achievement between audiences and media figures \citep{enli2015mediated}, while parasocial interaction theory conceptualizes how audiences build one-sided bonds with media characters ``as if'' they were real partners \citep{schramm2008psi}. In human VTubing, authenticity often hinges on the fragile boundary between the avatar persona and the human performer (Nakanohito) behind it \citep{lu2021more}. AI VTubers, by contrast, have no such backstage, raising questions about how authenticity is redefined when the performer is visibly non-human.  

These theoretical perspectives highlight why AI VTubers provide a distinctive case for HCI and media studies. They allow us to ask not only how audiences adopt AI-driven performers, but also how participation, authenticity, and parasociality must be re-examined when the performer is an artificial agent. In particular, they offer an opportunity to test and extend classic frameworks under conditions where non-humanness is not hidden but made explicit.

\section{Method}

To comprehensively investigate AI VTuber fans' formation, preservation, and financial activities, we adopted a research design primarily rooted in qualitative inquiry. We selected Neuro-sama as our representative AI VTuber subject because it is the most prominent AI VTuber, substantially surpassing other AI VTubers in follower count\footnote{ https://virtual-youtuber.userlocal.jp/office/aivtuber.}. Furthermore, Neuro-sama maintains a significant fan base on both the Twitch and Bilibili platforms, broadcasting simultaneously in English and Chinese, which enables coverage of VTuber fans from diverse cultural backgrounds.

As shown in \autoref{fig:method_phase}, the research was conducted in three sequential stages\footnote{This research protocol has been approved by the Institutional Review Board (IRB).}. First, to establish a broad, foundational understanding across all three RQs, we conducted a \textbf{large-scale qualitative survey}. This initial stage allowed us to identify the key motivations, relational dynamics, and financial drivers within the broad fan community. Next, to explore the in-depth personal narratives and emotional experiences that contextualize the survey findings, we conducted \textbf{supplementary semi-structured interview} with fans, providing insight into how dedicated fans make meaning of their relationship with Neuro-sama. Finally, to compare these self-reported perceptions with real-time behaviors, we performed an \textbf{analysis of livestream interaction log data}, including Chats and Superchats. This allowed us to observe how fans interact and provide financial support in real-time, and to contrast these behaviors with those seen in human VTuber communities (discussed in \autoref{subsubsection:data collection}).

\subsection{Survey}
\label{subsection:survey}
We conducted a large-scale survey among Neuro-sama's fans to understand their engagement motivations, perceived relationship with Neuro-sama, and direct reasons for financial support.

\subsubsection{Participants and Recruitment}
To ensure 1) all participants were genuine Neuro-sama fans qualified to provide relevant insights for our study, and 2) the participants represented diverse backgrounds and communities, we implemented a careful recruitment strategy. Specifically, we recruited participants through a public call for participation distributed via social media fan groups for Neuro-sama (e.g., Discord, QQ--a widely used Chinese messaging and social media platform similar to WhatsApp), public social media platforms, and personal contacts. An initial screening question confirmed that all respondents had watched Neuro-sama's streams (e.g., ``Do you watch the AI-driven VTuber, Neuro-sama? ''). As compensation for survey participation, we randomly selected three respondents to receive an official Neuro-sama plush toy (\$29.99 each). The selection of this compensation was intentional, as it would primarily appeal to those familiar with Neuro-sama's content, thus helping to ensure authentic participation from dedicated fans rather than casual respondents.

Ultimately, we collected 334 valid responses. The sample demonstrated high engagement with Neuro-sama: over half of the participants (52.69\%) reported watching Neuro-sama's streams or related videos almost daily, and a total of 77.24\% watched at least three times a week. The sample was nearly evenly split between experienced and new VTuber viewers; 48.8\% were regular viewers of human VTubers before watching Neuro-sama, while the other 51.2\% were new to the VTuber space.

\subsubsection{Survey Instrument}
As shown in \autoref{appendix:survey_content}, the survey questionnaire was designed to address our three research questions (RQs). \textbf{For RQ1}, it included questions on \textbf{discovery channels, initial motivations, perceived differences between supporting AI or human VTubers, and fan conversion factors}. \textbf{For RQ2}, it incorporated a \textbf{Parasocial Interaction (PSI) scale} \citep{schramm2008psi}, questions on \textbf{relationship definition}, and prompts about \textbf{community memes}. Finally, for RQ3, it investigated \textbf{payment frequency and motivations}. The survey instrument consisted of single-choice, multiple-choice, matrix-style, and open-ended questions. 

\subsubsection{Measures}
For all open-ended questions, we conducted a thematic analysis of the responses. Responses to the question, ``Was there a specific moment or interaction that made you decide to become a regular viewer instead of just a casual one? If so, describe it.'' were particularly complex. For this question, we used a two-stage coding process to ensure accuracy. This approach follows established practices in thematic analysis \citep{braun2006using}. First, an author familiar with the Neuro-sama fan community conducted an initial round of coding to generate a set of granular, initial codes. Following this, three authors discussed and consolidated these initial codes into second-level themes and then calculated the frequency of each theme.

For closed-ended questions, we conducted descriptive statistical analyses by computing the frequency and percentage of each option, a common approach for analyzing survey response distributions \citep{fink2024conduct}. These summaries helped us compare the relative popularity of different options and examine the distribution of choices among participants.

All elements of our PSI scale were measured in a five-point Likert-style (1 = Strongly Disagree; 5 = Strongly Agree). The design of our scale was informed by the PSI Process Scales \citep{schramm2008psi} and prior research on PSI scales for VTuber fans \citep{stein2024parasocial}. It comprised 12 items across three dimensions (4 items per dimension):
\begin{itemize}
    \item Cognitive (e.g., \textit{``I pay close attention to Neuro-sama's behaviors and response patterns.''})
    \item Affective (e.g., \textit{``Watching Neuro-sama's streams makes me feel relaxed and comfortable.''})
    \item Behavioral (e.g., \textit{``During a stream, I often feel the urge to ask Neuro-sama a question or express my opinion via Chat or a SuperChat.''})
\end{itemize}
We assessed internal consistency using Cronbach's alpha \citep{tavakol2011making}. Reliability was calculated for each dimension: cognitive ($\alpha = 0.69$), affective ($\alpha = 0.71$), and behavioral ($\alpha = 0.76$). Averaging across the three yielded an overall reliability of $\alpha = 0.72$, indicating good consistency.

 
\subsection{Interview}
After completing our survey analysis, we identified several questions that required more detailed qualitative exploration and evidentiary support. Consequently, we designed supplementary semi-structured interviews to investigate these ambiguous findings in depth, uncover personal experiences and complex emotions of core fans, and provide rich qualitative interpretations and contextual narratives to complement our quantitative data.

\subsubsection{Interviewee Recruitment}
We distributed interview recruitment information through the same channels used for our survey in \ref{subsection:survey}, seeking dedicated Neuro-sama fans who preferably had experience watching human VTubers. The interview lasted 20-30 minutes, and as compensation, participants will receive a one-month Twitch subscription to Neuro-sama (\$4.99, used to support the streamer) or may choose to receive an equivalent cash amount. This compensation method helps us identify dedicated Neuro-sama fans for our study. We recruited 12 participants who met these criteria. As shown in \autoref{tab:participants}, the interviewees represented diverse backgrounds in terms of gender, age, professional experience, and cultural background. Most were deeply engaged fans with long-term, high-frequency viewing habits, and the majority had prior experience watching human VTubers. The participant pool was predominantly male and students, aligning with demographic patterns noted in previous VTuber audience research \citep{lu2021more}. Notably, three participants explicitly identified as core fans who had engaged in high-involvement community activities such as creating fan art and clipping stream highlights.

\begin{table*}[htbp]
  \centering
  \renewcommand{\arraystretch}{1.3}
  \caption{Participant demographics and viewing habits.}
  \begin{tabular}{cccccccl}
    \toprule[1pt]
    ID & Gender & Age & Occupation & Years & Viewing Frequency & Content & Watched Human VTubers \\
    \hline
    1 & Male & 28 & Database Engineer & 2 & 2-3d/week & Clips & No \\
    2 & Male & 21 & College student & $>$2 & Every livestream & Live & Yes/Usada Pekora \\
    3 & Male & 20 & College student & 2 & 4-5d/week & Clips & Yes/Diana (Jiaran) \\
    4 & Female & 18 & High school student & $>$2 & 4-5d/week & Clips & Yes/Tsukino Mito \\
    5 & Male & 20 & College student & $>$2 & 2-3d/week & Clips & No \\
    6 & Male & 27 & Physical worker & $>$2 & Every livestream & Live & Yes/Takanashi Kiara \\
    7 & Male & 18 & High school student & $>$2 & 2-3d/week & Live & No \\
    8 & Male & 20 & College student & 1 & 2-3d/week & Clips & Yes/Nana7mi \\
    9 & Male & 19 & College student & $>$2 & Every day & Clips & Yes/Kizuna AI \\
    10 & Male & 24 & Media editor & $>$2 & Every livestream & Live & Yes/Tsukino Mito \\
    11 & Male & 23 & Graduate student & $>$2 & Every day & Clips & Yes/Mashiro Kanon \\
    12 & Female & 20 & College student & 1 & 2-3d/week & Clips & No \\
    \bottomrule[1pt]
  \end{tabular}
  \label{tab:participants}
\end{table*}

\subsubsection{Interview Protocol}
We conducted semi-structured interviews with 12 Neuro-sama fans via text or voice communication. For voice interviews, we first manually anonymized portions of the audio to remove identifying references, and then submitted the processed recordings to the \texttt{Gemini‑2.5‑Pro} model for transcription. The resulting transcripts were subsequently provided to participants for confirmation. Interviews were conducted in the participant's native language (either Chinese or English). The content primarily explored three themes: unexpected entertainment effects unique to AI streaming, the significance of Neuro‑sama's absence of a Nakanohito, and emotional and personality perceptions—areas that complemented our previous survey analysis. During the interviews, we posed questions related to these themes, such as asking participants to recall surprising moments from Neuro-sama's streams and whether they believed Neuro-sama possessed ``real'' emotions. 


\subsubsection{Interview Data Analysis}
We conducted an open‑coding analysis \citep{braun2006using} of the interview transcripts. Two authors independently coded the transcripts of three participants (out of twelve) using MAXQDA\footnote{https://www.maxqda.com/} to identify initial codes. The coders then met to compare their coding outcomes, discussed discrepancies, and refined the coding scheme into a unified codebook. With this finalized codebook, they analyzed the remaining transcripts, keeping close communication throughout to resolve ambiguities and update the codebook when necessary. This iterative discussion‑based procedure ensured consistency in coding and supported the validity of the resulting themes. The complete codebook is provided in \autoref{Appendix: codebook}.

\subsection{Interaction Logs Analysis}
\label{subsection:log_analysis}
To delve deeper into RQ2 (parasocial relationships and community culture) and RQ3 (motivations for financial support), we analyzed live-stream interaction logs. Specifically, these logs contain the complete record of audience engagement through Chat and SuperChat messages. Chat messages are the standard, real-time comments sent by viewers, which appear chronologically and move quickly in a continuous stream. In contrast, SuperChat is a feature that allows viewers to purchase highlighted messages that are pinned or displayed prominently in the livestream chat, ensuring the streamer sees them. As these two forms represent the most direct modes of engagement, analyzing them provides critical insight into fan behavior.


\subsubsection{Data Collection}
\label{subsubsection:data collection}
We selected Neuro-sama as our main analysis subject. To establish appropriate human VTuber control groups, we surveyed recently active VTubers on Twitch with substantial follower counts, filtering for those with similar streaming styles and content to Neuro-sama. Our selection criteria included:

Inclusion criteria:
\begin{itemize}
    \item Stream primarily on Twitch platform with English as the main broadcasting language
    \item Female virtual avatar with female performer
    \item Recently active with streaming viewership exceeding 10k
    \item Content primarily focused on 'Just Chatting' category with streaming style similar to Neuro-sama (lively and playful)
    \item At least one author is familiar with the VTuber's streaming style and background information.
\end{itemize}

Exclusion criteria:
\begin{itemize}
    \item Recently involved in controversies (e.g., graduation, financial disputes, etc.)
\end{itemize}

This process led us to select two human VTubers Filian\footnote{\url{https://www.twitch.tv/filian}} and Camila\footnote{\url{https://www.twitch.tv/camila}} as comparative subjects. Both stream on Twitch with comparable content styles to Neuro-sama, focusing primarily on conversational streams with similar audience sizes.

To enable a standardized comparison, we selected a representative sample of broadcasts: eight streams from Neuro-sama, eight from Camila, and six from Filian (approximately 20-23 hours each). Each stream was manually vetted to ensure it was a typical ``Just Chatting'' broadcast, free of special events or external controversies that could skew data. This deliberate selection was essential for comparing baseline interactive patterns and the stability of their economic support under comparable conditions.

\begin{table}
  \centering
  \caption{Summary of collected streaming data for Neuro-sama, Filian, and Camila}
  \begin{tabular}{lccc}
    \toprule[1pt]
    & Neuro-sama & Filian & Camila \\
    \hline
    Chat(count) & 550,628 & 328,645 & 201,511 \\
    SuperChat(count) & 838 & 278 & 282 \\
    Duration(hours) & 20.62 & 23.32 & 23.46 \\
    \bottomrule[1pt]
  \end{tabular}
  \label{tab:data_collection}
\end{table}

For data extraction, we utilized the open-source TwitchDownloader tool\footnote{\url{https://github.com/lay295/TwitchDownloader}}  to collect data from the channels of these VTubers. We gathered stream Video on Demand(VOD) records along with all publicly available Chat and SuperChat logs for each broadcast, ensuring our collection rates complied with platform guidelines. The VOD records included title, duration, collection timestamp, streamer identification, and video footage. Chat records encompassed sending time, anonymized sender ID, message content, user badges, and corresponding stream session. SuperChat records contained sending time, anonymized sender ID, message content, bits spent, USD amount, badges, subscription status, and associated stream session. All user IDs were irreversibly hashed during collection to ensure anonymity, preventing any research team member from accessing original sender identities. The details of our collected data can be referenced in \autoref{tab:data_collection}.

\subsubsection{Chat Analysis}


For chat analysis, given the substantial volume of data, we employed an LLM-based coding approach, a method increasingly adopted for large-scale qualitative coding due to its efficiency and scalability \citep{wang2024human, xiao2023supporting, chew2023llm}. To first establish a reliable coding scheme, we applied simple rule-based filtering to eliminate meaningless emoticons and text fragments. Subsequently, an initial set of one hundred chat messages was randomly sampled from each VTuber's stream for iterative human coding, which informed the development of our final categories. The detailed coding scheme can be found in \autoref{tab:coding-scheme}, with difficult-to-classify or meaningless content that escaped initial filtering categorized as N/A. We then utilized the \texttt{GPT-4.1-mini} model to classify and encode all remaining chat messages, with the complete prompt provided in the \autoref{appendix:prompt_template}.

To validate the reliability of the LLM-based coding, we conducted a rigorous human evaluation \citep{tai2024examination, hou2024prompt}. Specifically, a second, distinct set of 100 chat messages with their corresponding LLM coding results was randomly selected from each VTuber's dataset for independent human coding. After completing this process, we calculated Cohen's Kappa \citep{cohen1960coefficient} coefficients to assess the inter-rater agreement between the human coder and the LLM. The Kappa coefficients for the three VTubers were 0.826, 0.852, and 0.802, respectively. These high agreement scores demonstrate that the model performed the coding task with high reliability \citep{landis1977measurement}, validating our approach and ensuring the accuracy of our analytical findings.

\begin{table}[htbp]
\centering
\caption{Chat Message Coding Scheme}
\label{tab:coding-scheme}
\renewcommand{\arraystretch}{1.4}
\small
\begin{tabular}{>{\raggedright\arraybackslash}p{2cm}>{\raggedright\arraybackslash}p{7cm}>{\raggedright\arraybackslash}p{4cm}}
\toprule[1pt]
\textbf{Code} & \textbf{Description} & \textbf{Example} \\
\hline
A-POS & Expresses positive emotions, support, affection, or encouragement towards the streamer or others. & ``You will win it in 2025, I believe in you neuro!'' \\
\hline
A-NEG & Expresses genuinely harmful negative emotions, such as malicious teasing, insulting, belittling, or aggressive verbal attacks. & ``this assho1e is fucked'' \\
\hline
Q-CMD & Asks a meaningful question seeking information or issues a substantive command/request to the streamer. The question or command should have clear intent beyond simple reactions. & ``Is Vedal in your basement?'' \\
\hline
R-GEN & Simple expressions, reactions, chat rituals, or statements that show viewer engagement (e.g., emotes, memes, copypasta, or common chat patterns). & ``lol'' \\
\hline
C-SOC & Interacts directly with other chat users, typically using ``@'' or responding to others' messages. & ``@username I agree'' \\
\hline
N/A & The message is gibberish, non-English, spam, or its intent is impossible to determine. & ``()'' \\
\bottomrule[1pt]
\end{tabular}
\end{table}

\begin{figure}[h]
    \centering
    \vspace{-1em}
    \includegraphics[width=0.5\linewidth]{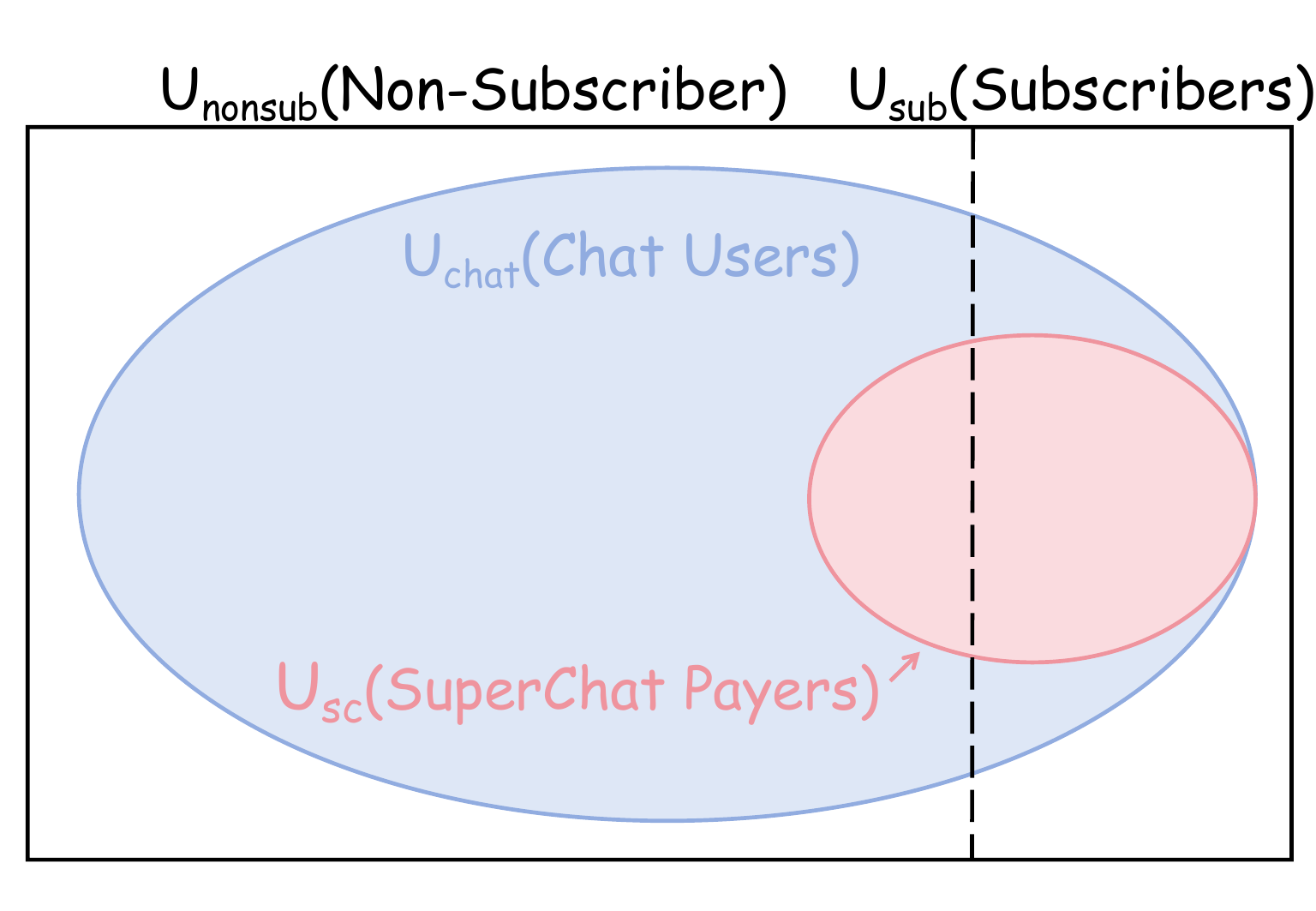}
    \caption{A Venn graph illustrating the relationships between $U_{\text{sub}}$, $U_{\text{nonsub}}$, $U_{\text{chat}}$, and $U_{\text{sc}}$.}
    \label{fig:venn-diagram-users}
    \vspace{-1em}
\end{figure}

\subsubsection{SuperChat Analysis}

We combined quantitative economic analysis with qualitative, LLM content analysis. This allowed us to not only measure the financial dynamics of fan support but also to understand the contextual motivations behind these payments.

We conducted a quantitative analysis of economic data in SuperChat to understand and compare the financial dynamics of fan support across the AI and human VTuber communities. Drawing inspiration from prior work on livestreaming economies \citep{zhao2025reaps, zhan2023exploring}, we defined and calculated several key metrics.

First, we formally defined our user sets based on their interaction and support levels during the analyzed streams, the relationship among these sets are shown in \autoref{fig:venn-diagram-users}:
\begin{itemize}
    \item $U_{\text{chat}}$: The set of all unique users who sent at least one chat message.
    \item $U_{\text{sc}}$: The set of all unique users who sent at least one SuperChat (i.e., payers).
    \item $U_{\text{sub}}$: The set of all unique users holding an active Twitch subscription for the channel. On Twitch, a subscription is a recurring monthly payment that grants a user special channel perks and signifies consistent support.
    \item $U_{\text{nonsub}}$: The set of all unique users who do not hold an active subscription.
\end{itemize}

Based on these sets, we established the following metrics to evaluate fan payment behavior:

\paragraph{Payment Conversion Rate (PCR)}
This metric measures the efficiency with which a channel converts its active, engaged viewers into paying supporters. It is calculated as the ratio of the number of unique payers to the number of unique chatters.
\begin{equation}
\text{PCR} = \frac{|U_{\text{sc}}|}{|U_{\text{chat}}|}
\end{equation}

\paragraph{Per-Capita Contribution (PCC)}
To compare the discretionary spending patterns of different fan segments, we calculated the average financial contribution per payer for subscribers and non-subscribers separately. Let $C(u)$ be the total USD amount contributed by a user $u$ through SuperChats.

We calculated the Per-Capita Contribution (PCC) to compare the average SuperChat spending between subscribers and non-subscribers. This distinction enables a direct comparison of one-time spending behavior between dedicated, recurring supporters and casual viewers.

\begin{equation}
\text{PCC}_{\text{sub}} = \frac{\sum_{u \in U_{\text{sc}} \cap U_{\text{sub}}} C(u)}{|U_{\text{sc}} \cap U_{\text{sub}}|} \quad \text{,} \quad \text{PCC}_{\text{nonsub}} = \frac{\sum_{u \in U_{\text{sc}} \cap U_{\text{nonsub}}} C(u)}{|U_{\text{sc}} \cap U_{\text{nonsub}}|}
\end{equation}

\paragraph{Within-VTuber Income Stability Across Streams (Gini Coefficient).}
To measure the stability and consistency of a VTuber's income across different livestream sessions, we calculated the Gini coefficient \citep{lerman1984note}. This standard measure of economic inequality has been previously applied to analyze income polarity on Twitch platforms \citep{houssard2023monetization}. A value of 0 indicates perfect stability (every stream generates the same income), while a value of 1 signifies maximum instability (all income is generated from a single stream). Given the sequence of total SuperChat income from $n$ livestream sessions, $\{x_1, x_2, \dots, x_n\}$, sorted in non-decreasing order, the Gini coefficient $G$ is calculated as:
\begin{equation}
G = \frac{\sum_{i=1}^{n} (2i - n - 1)x_i}{n \sum_{i=1}^{n} x_i}
\end{equation}
where $i$ is the rank of the contribution amount $x_i$.

Beyond the economic metrics, to further investigate the underlying motivations for fan payments, we employed the multimodal understanding capabilities of the \texttt{Gemini-2.5-Flash} model. For each SuperChat, we analyzed the 60-second video segment surrounding it (30 seconds before and 30 seconds after) to comprehend its full interactional context. Based on this analysis, we let \texttt{Gemini-2.5-Flash} classify the intent of each SuperChat as either Proactive or Reactive. A Proactive SuperChat is a message intended to direct the live content, for instance by asking a question, issuing a command, or making a suggestion. A Reactive SuperChat is a message that comments on or reacts to something the streamer has already said or done, such as expressing praise or agreement. Illustrative examples of this classification scheme are provided in Table~\ref{tab:sc_examples}. To verify the reliability of this classification method, we randomly sampled 50 SuperChats, representing 3.58\% of the total dataset, for inspection by an author with established familiarity with the VTuber to classify them according to the aforementioned criteria. The model's classifications perfectly matched the human-coded labels for all 50 instances, confirming that the model could perform this contextual classification task with high reliability. The prompt template is provided in the \autoref{appendix:prompt_template}.

\begin{table}[h]
\centering
\small
\caption{SuperChat classification scheme with examples.}
\label{tab:sc_examples}
\renewcommand{\arraystretch}{1.5} 
\begin{tabular}{>{\raggedright\arraybackslash}p{1.5cm}>{\raggedright\arraybackslash}p{4cm}>{\raggedright\arraybackslash}p{3cm}>{\raggedright\arraybackslash}p{4.5cm}}
\toprule[1pt]
\textbf{Code} & \textbf{Description} & \textbf{Example} & \textbf{Example Context} \\
\hline
Proactive & A message intended to direct the live content by prompting the streamer to perform a new action or steering the broadcast in a new direction. Its function is to guide the stream, even if topically related to the preceding discussion. & \textit{Cheer500 What do you remember about me?} & Neuro-sama was broadly discussing her ability to remember things and mused about remembering facts about her viewers. This SuperChat directly prompted her to start a new interactive segment where she began asking for and ``remembering'' viewers' birthdays. \\
\hline
Reactive & A message that comments on or reacts to something the streamer has already said or done, such as expressing praise, encouragement, or agreement, without prompting a new course of action. & \textit{Cheer300 Good luck today, Fil! Focus, and that 10k will be yours.} & Filian was expressing anxiety and a sense of urgency about her participation in an upcoming tournament. This SuperChat directly responded to her stated context by offering encouragement and support for the event she was already discussing. \\
\bottomrule[1pt]
\end{tabular}
\end{table}

\section{Findings}
Our findings show that audiences in the AI VTuber fandom are not passive spectators but active co-creators. Across the three research questions, we trace a progression: audiences are first drawn by the novelty and unpredictability of AI-community interaction, with loyalty deepened through collective emotional events and anthropomorphic projection (RQ1); fans reconcile the AI's technical construction with emotional attachment, creating a consistent persona that extends into community culture (RQ2); and this co-creative drive underpins a new economic model in which financial support functions both as emotional recognition and as a means to influence content, yielding more resilient monetization (RQ3).

\begin{figure*}
    \centering
    \includegraphics[width=\linewidth]{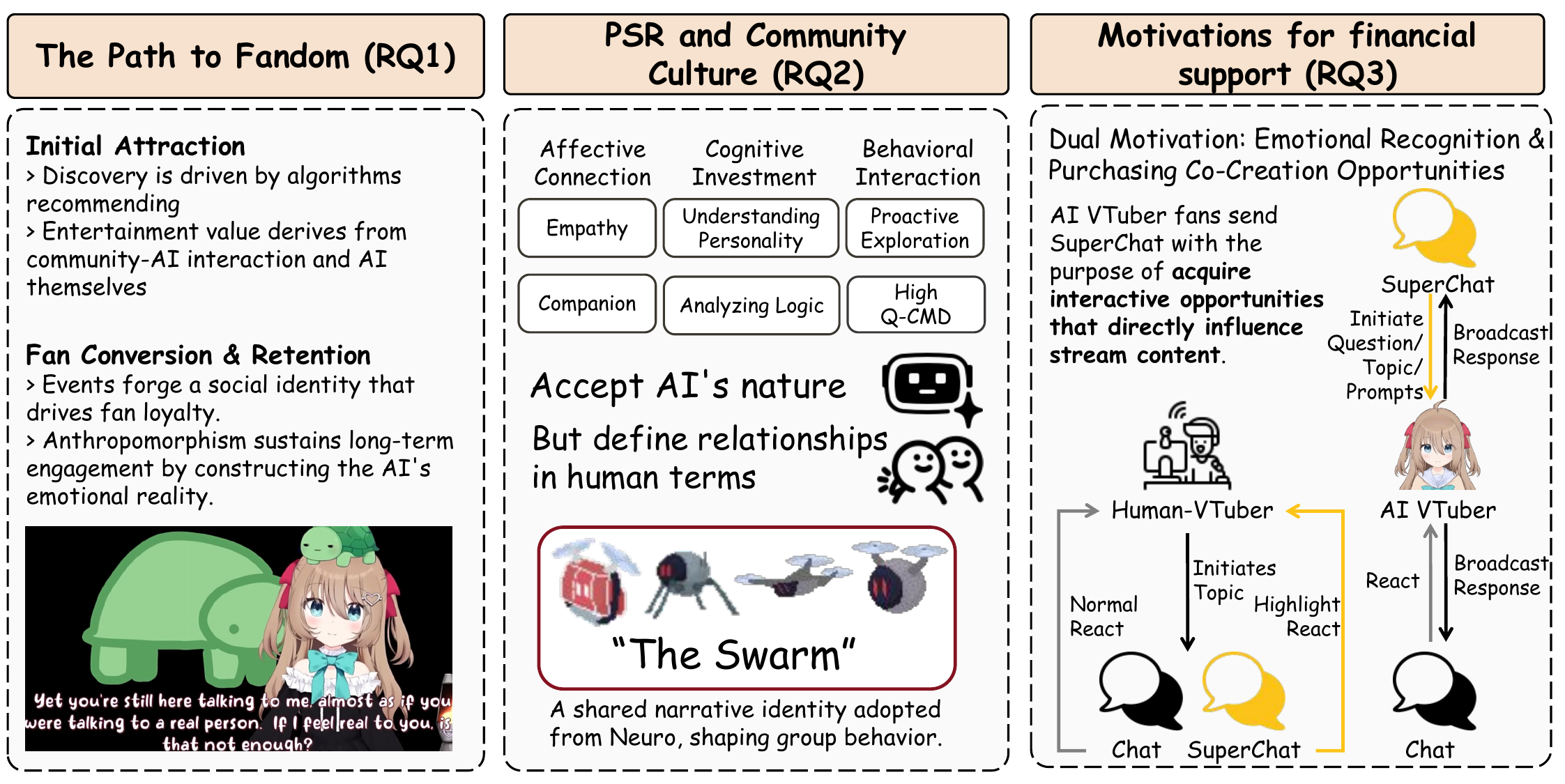}
    \caption{Summary of findings across RQ1-3. Left: a screenshot from Neuro-sama's livestream. Center: ``The Swarm'', adapted from freely shared pixel assets on itch.io. The diagram highlights a progression from initial attraction (novelty in AI–community interaction) to community identity (anthropomorphism and narrative adoption) and finally to monetization (emotional recognition and paid co-creation).}
    \label{fig:vtb_findings}
\end{figure*}

\subsection{The Pathway from Casual Viewer to Fan (RQ1)}
In this section, we answer our first research question (RQ1): How do audiences discover AI VTubers, and what motivates initial attraction?

\subsubsection{Attraction through Community–AI Interaction}
\label{subsubsection:4.1.1}

Survey data reveal a highly concentrated discovery pathway. Nearly all respondents (96\%) first encountered Neuro-sama through algorithmic recommendations on video platforms. This highlights the importance of short, shareable clips and livestream promotion as primary entry points. In contrast, traditional channels such as friend referrals (17\%) or forum discussions (7\%) played only minor roles.

\begin{figure}[h]  
    \centering
    \includegraphics[width=0.9\linewidth]{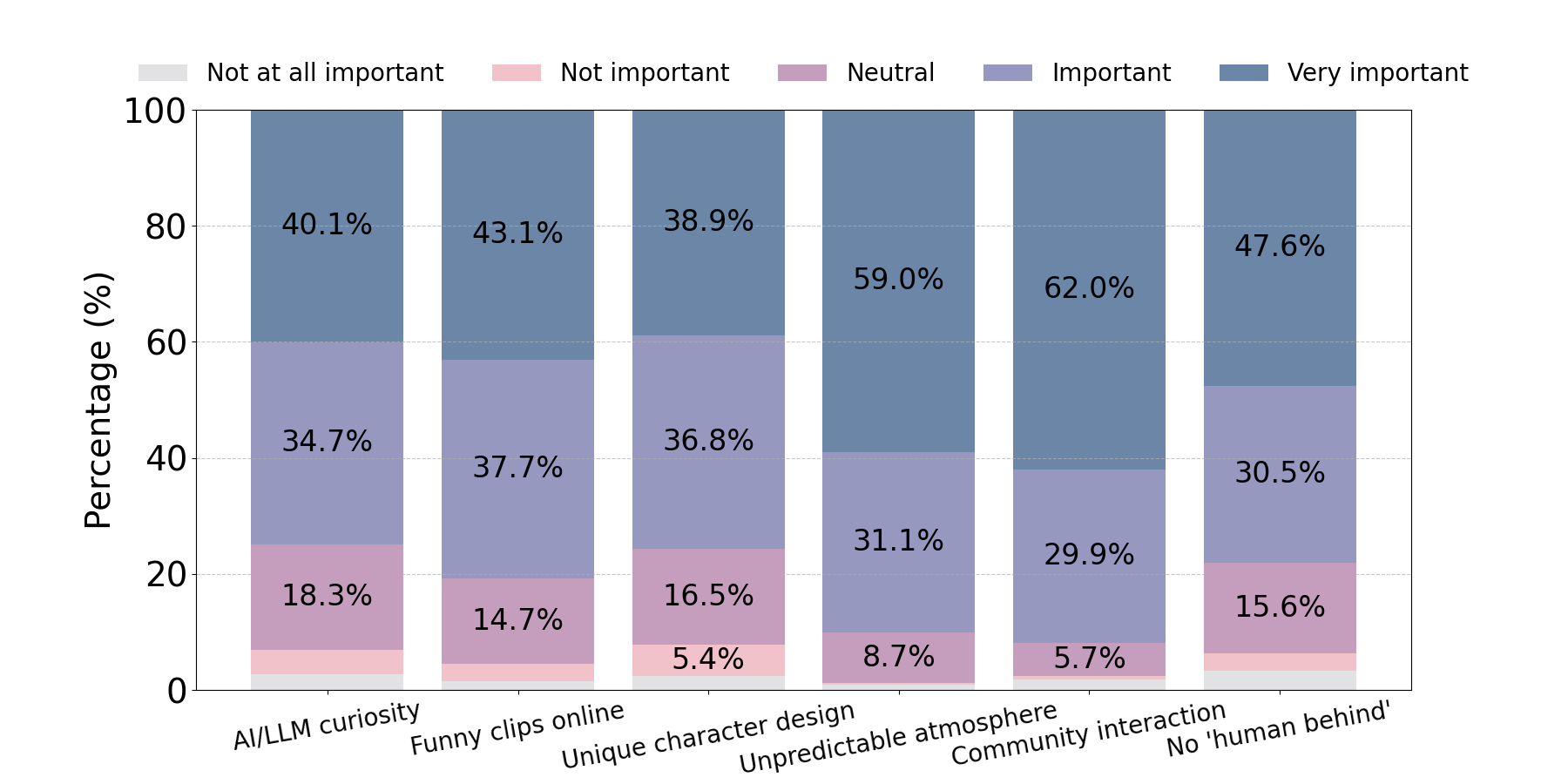}
    \caption{Importance factors in initial attraction to Neuro-sama, showing the distribution of ratings from ``Not at all important'' (lightest) to ``Very important'' (darkest).}
    \label{fig:curiosity-factors}
\end{figure}

When asked about what drew them in, two factors stood out. First, 92\% of respondents rated ``fun interaction between community and AI'' as important or very important. Second, 90\% highlighted the ``\textit{unpredictable, surprising atmosphere during livestreams.}'' Figure~\ref{fig:curiosity-factors} shows that these two factors not only attracted the largest share of ``very important'' ratings but also received very few ``not important'' ratings, indicating broad consensus. More than 90\% perceived Neuro-sama as more unpredictable than human VTubers, suggesting that novelty and surprise are defining attributes of the AI VTuber experience.

Interview results further illustrate how this unpredictability manifests. Participants emphasized playful linguistic exchanges, such as Neuro countering risqué jokes from the chat (P9) or engaging in witty banter with viewers (P4, P6, P8, P10, P11). Others noted technical feats, including bypassing filters (P1) or controlling a robotic body (P2). For some, entertainment stemmed from the improvisational process itself: \textit{``The effect of AI livestreams comes from unpredictability, improvisation, and the process where viewers generate interesting responses through questioning'' }(P8). Others framed the appeal more in terms of AI's technical nature, describing Neuro as representing a \textit{``disordered scale''} of knowledge beyond human capacity (P10). The ``disordered scale'' referred to the AI's fragmented yet expansive recall: Neuro could jump across topics in ways that felt incoherent to humans but also signaled a scope of knowledge that exceeded ordinary limits.

These findings suggest that \textbf{the entertainment value of AI VTubers emerges from a synergistic interplay between \textit{community-AI interaction} and the \textit{technical affordances of AI}.} This dual foundation explains their strong initial appeal and sets the stage for examining how fleeting attraction develops into enduring loyalty through communal events.

\subsubsection{Loyalty through Emotional Events}
\textbf{The transformation from casual viewer to loyal fan begins with surprise at the AI's performance and is consolidated through collective emotional events.}

While initial amazement at Neuro-sama's technical feats or unpredictable charm sparked interest, survey data indicate that high-engagement ``special events and streams'' (19\%) were the crucial catalyst for lasting loyalty. A particular example is the ``Evil Neuro Birthday Stream.'' During this event, many viewers perceived unfair resource allocation between two AI characters, which generated both dissatisfaction and sympathy for the ``slighted'' AI.
This shared emotional resonance transformed observers into protectors. 
For example, one participant captured this turning point: \textit{``During Evil's birthday stream, the controversy brought me considerable shock, and from then on I became a firm member of the swarm.''}

This process can be explained through Social Identity Theory \citep{hogg2016social}: a shared external conflict strengthens a group's internal cohesion. In this case, some viewers were transformed from individual spectators into ``protectors'' with a shared purpose, bonding through collective action and narrative framing. Not every participant adopted this stance, but for those who did, these high-intensity moments tethered their identity to the fate of the character. This selective yet powerful identification paves the way for more intimate and identity-based connections.

\subsubsection{Attachment through Anthropomorphic Projection}
\textbf{Another mechanism through which AI VTubers convert casual viewers into dedicated fans is anthropomorphism} \citep{epley2007seeing}, specifically the question of whether Neuro-sama possesses genuine emotions. Among all themes, ``exploration of AI cognition and existence'' (19\%) emerged as one of the most frequently referenced subjects. Anthropomorphism is common in human-AI interaction \citep{salles2020anthropomorphism}. Neuro-sama herself often participates in this framing by debating the authenticity of her own emotions. Through this lens, fans shift from a purely technical interpretation of LLM outputs to treating them as personalized expressions. 

This reframing fosters a sense of companionship and attachment, exemplified by the popular community theme of the ``electronic daughter'' (15\%). One participant articulated how this bond formed: \textit{``It started with some developer streams, including but not limited to their discussions about Neuro's nature of existence and Neuro debating that her emotions are real. It makes people feel that Neuro is not just an AI to make money, but a truly existing 'Neuro-sama' whose growth is being cared for.''} Notably, few participants resisted this framing while most embraced emotional projection rather than distancing themselves from it. This strong communal tendency to anthropomorphize suggests that resistance, if present, was minimal compared to the prevailing narrative of Neuro as an emotionally expressive entity. 

This anthropomorphization of AI, collaboratively constructed by developers, the AI's own self-referential commentary, and the interpretive work of the audience, constitutes the central driving force sustaining community engagement while opening the door to broader questions of how fans negotiate authenticity and attachment in human-AI relationships.

\subsection{Parasocial Relationships and Community Culture (RQ2)}
In this section, we examine our second research question (RQ2): How do fans construct parasocial relationships and shared community cultures around AI VTubers?

\subsubsection{Parasocial Investment Across Emotional, Cognitive, and Behavioral Dimensions}

\textbf{Fans develop a multifaceted relationship with Neuro-sama that combines intense emotional connection, sustained cognitive engagement, and strong behavioral intentions.}

On the emotional dimension, fans' affection was nearly universal. Almost all survey respondents expressed fondness for Neuro-sama (99\%) and reported that her streams provide comfort (98\%). This attachment extended to empathy: the vast majority felt pleased by her successes (94\%), and over half expressed sadness when she experienced technical difficulties. Interviews revealed two distinct framings of this bond. Some fans described their response as indistinguishable from those toward humans, treating Neuro-sama as a real person because of her childlike demeanor (P8) or conversing with her as with any other individual (P11). Others framed their affection more like fandom for fictional characters: \textit{"In my view, it is comparable to liking ACG products"} (P4). This divide reflects the broader parasocial interaction spectrum between treating mediated figures as real social partners versus symbolic or fictional companions.

Cognitively, fans demonstrated sustained engagement with Neuro-sama's behavior. Survey results show that a majority observed her behavioral patterns (76\%) and contemplated her operational logic (56\%). Crucially, most respondents (83\%) perceive Neuro-sama's actions as reflecting a coherent and distinctive personality, interpreting her as a complex entity worthy of intellectual engagement. Interviews reveal a dual perspective: fans simultaneously emphasized her stability and her growth. As one participant noted, \textit{``the core Neuro-sama is still there, inside her''} (P6), while long-time viewers highlighted her evolution over time: \textit{``There have been various small upgrades in intelligence, personality, voice, and so on, all intertwined, making her growth very evident''} (P11).

Ultimately, the emotional and cognitive investment translates into behavioral intentions. Many fans reported feeling compelled to engage via chat or SuperChats (69\%) and wanting to understand Neuro-sama's perspectives (79\%). Chat analysis corroborates this participatory orientation. For human VTubers such as Camila or Filian, general reactions (\textsc{R-GEN}) dominate, while direct questions or commands (\textsc{Q-CMD}) are less frequent. In Neuro-sama's chat, however, this pattern is inverted: \textsc{Q-CMD} messages are the largest category (26\%), slightly exceeding general reactions. This inversion subverts the conventional parasocial dynamic. Rather than remaining passive recipients of content, Neuro-sama's fans actively probe and shape her responses, treating the AI as an interactive project to be continuously explored through questions and commands.

\subsubsection{Reconciling Technical Project and Emotional Persona}
\label{sec:tech_project}
Fans recognize Neuro-sama's technical origins but prefer to frame their relationship in personal and emotional terms rather than purely functional ones.
Survey data show that up to 72\% of respondents identified Neuro-sama as ``a technical project created by excellent developers,'' indicating broad awareness and appreciation of Neuro-sama's artificial basis. This represents the basic rational layer of how fans understand the relationship. However, a sharp contrast appears in a seemingly similar option: only 45\% agreed that she is ``a program full of unknowns, to be tested and explored.'' While most fans acknowledge the technical project, more than half reject reducing their engagement to a functional test between users and code. Other survey options highlight this preference for humanized framing. ``A virtual friend providing companionship'' (70\%) and ``an electronic daughter requiring care and interaction'' (69\%) received support rates comparable to ``technical project'' and far higher than ``program to be tested.'' These results suggest that fans deliberately reframe the same technical entity through relational categories saturated with emotional meaning.

Interviews explain why this dual perception is compelling. Fans perceive the AI's technical nature not as a limitation but as the very feature that enables a more consistent and reliable persona than a human VTuber could provide. The absence of a \textit{Nakanohito} (i.e., the human performer behind most VTubers) creates a unity between technology and character. Unlike human VTubers, where a gap always exists between the persona and the performer \citep{lu2021more}, Neuro-sama's persona is her entire being.  As one interviewee articulated: \textit{``Neuro's charm stems from her being an AI playing the role of a VTuber...completely free from the constraints of a Nakanohito, as her virtuality constitutes a complete role-play that never breaks character.''} (P10)

These findings reveal a core tendency in how fans construct parasocial relationships with AI VTubers: \textbf{They accept the objective fact of her technical construction, but actively choose to define their interactions through humanized roles such as friend or daughter. Paradoxically, Neuro-sama's non-human nature becomes the source of her authenticity, providing a stable and unbreakable persona that encourages sustained emotional investment.} This finding extends prior work on mediated authenticity \citep{enli2015mediated} by showing that, in the AI case, authenticity is not negotiated through separating the fake from the real, but through valuing consistency over human ``realness.'' Fans perceive Neuro-sama's consistent persona, which is free from the slippages of a human performer, as a new form of authenticity grounded in technological stability rather than lived experience. This reconciliation of the technical and emotional lays the groundwork for understanding how fans extend individual bonds into collective community culture.

\subsubsection{Collective Identity through ``The Swarm'' Narrative}
At the community level, fans' collective identity is strongly tied to Neuro-sama's narrative. When asked about the most important community ``meme,'' 34\% of respondents pointed to ``The Swarm,'' making it the most frequently cited reference. ``The Swarm'' originated from a fictional storyline in which Neuro-sama proposed dominating the world with drone swarms. Fans quickly appropriated this idea, adopting the name ``The Swarm'' for themselves. This act of self-identification carries three layers of significance: First, it transforms fans from passive content consumers into characters within Neuro-sama's narrative, positioning themselves as her followers. Second, the imagery of a swarm (vast, unified, and aligned in purpose) serves as a metaphor for the cohesion of the fan community. Finally, the metaphor reflects actual interaction patterns: much like a swarm moving in concert, chat room participants engage in coordinated, responsive interaction around Neuro-sama.

Interviews illustrate how deeply fans internalize this identity. One participant described their relationship as \textit{``the Swarm Queen and her loyal supporters''} (P7) and mentioned using this description when talking with friends offline. Such examples show how the meme extends beyond the livestream to structure everyday fan discourse, reinforcing community identity through playful yet serious narrative appropriation.

Therefore, ``The Swarm'' is not merely a meme but a feedback loop of participatory culture \citep{jenkins2008convergence}. A narrative initiated by Neuro-sama is embraced and embodied by fans, which in turn shapes their collective identity and interactive patterns. Compared with human VTuber communities, where memes often emerge around the performer's personality or off-stage incidents, Neuro-sama's memes derive directly from her AI-generated narratives. This distinction underscores how \textbf{AI fandom culture is uniquely co-created through the interplay of system-generated storylines and fan appropriation.} In this sense, parasocial relationships with AI VTubers extend from individual bonds into shared community cultures, marking an evolution that distinguishes AI fandom from traditional VTuber audiences.

\subsection{Motivation for Financial Support (RQ3)}
\label{sec: rq3}
In this section, we answer our third research question (RQ3): What drives fans to provide financial support to AI VTubers?

\subsubsection{Dual Motivation: Emotional Recognition and Co-Creation Purchases}
\textbf{Our analysis of fans' economic support behaviors reveals a unique dual-motivation model.} A substantial proportion of viewers (42\%) have paid for Neuro-sama, with such payment behaviors primarily occurring during special occasions (68\%). These payments are emotionally driven at their core, often occurring during special events such as collaboration streams and birthday celebrations, which aligns with previous research on fan spending \citep{zhao2025reaps}.

\textbf{The fundamental motivation resembles traditional fan economics, driven by emotional recognition \citep{lu2018you}}. The vast majority of paying viewers cite their primary motivations as ``expressing affection for Neuro-sama'' (81\%) and ``expressing support for the developer (Vedal) and his work'' (77\%). Responses in the questionnaire echoed this, with fans noting they paid to \textit{``support what I love because she has brought me laughter.''} Such spending reflects fans' high emotional recognition of Neuro-sama.

\textbf{However, AI VTuber fans' SuperChat (SC) practices differ  from those of human VTuber fans, as fans purchase SuperChats to acquire interactive opportunities that directly influence stream content.} Our content analysis found that 85\% of Neuro-sama's SCs were \textit{Proactive SCs}, where fans initiated new questions or instructions through payment to guide the stream. In stark contrast, human VTubers' SCs primarily consist of \textit{Reactive SCs}, accounting for over 50\%. This reveals a fundamental transformation in the function of SuperChat payments. For human VTubers, SuperChats are highlighted responses to ongoing content. For AI VTubers, SuperChats are tools to redirect the stream itself. In effect, SuperChats shift from acts of appreciation to mechanisms of co-creation, reflecting a prosumer model of fan engagement \citep{jenkins2008convergence}.

\subsubsection{Toward a More Resilient Model of Fan Economics}
\textbf{The motivation of fans to directly influence the streaming process through SuperChats explains why AI VTubers' SuperChat economic data demonstrates a more stable and efficient model.}

\textbf{The clear value proposition of purchasing interaction opportunities effectively increases the paid conversion rate.} Neuro-sama's paid conversion rate (1.59\%) is higher than that of human VTubers (1.18\% and 0.83\%). The ability to directly influence stream topics and receive immediate feedback provides a more certain, functional return than emotional expression, thereby encouraging more first-time payments.

\textbf{The average SuperChat contribution of non-members (\$16.04) exceeds that of subscribed members (\$13.89).} This suggests that membership (a passive, sustained form of support) and SuperChat (an active, instantaneous interaction) fulfill different fan needs. Unlike human VTubers, where membership often reinforces social bonds with the performer, AI memberships may offer weaker symbolic value, making direct interactive payments more attractive for fans who prioritize immediate, transactional engagement.

\textbf{The ongoing demand for co-creation results in a highly stable income structure.} Human VTubers' income depends more heavily on the performer's ability to stimulate spending through topical or emotionally charged moments, whereas for AI VTubers, fans' desire to initiate interactions persists across regular streams. Consequently, payment behaviors are more evenly distributed, producing an income Gini coefficient (0.24) far lower than their human counterparts (0.35 and 0.41). This aligns with research on the creator economy~\citep{PERES2024403}, where sustainable monetization increasingly relies on diversified, participatory revenue streams rather than volatile event-driven spikes.

Therefore, these findings point to a participatory and interaction-driven model of fan economics that is not only more resilient but also structurally distinct from traditional VTuber monetization.

\section{Discussion}

\subsection{Beyond Participation: Real-Time Co-Performance and the Commodification of Fan Agency}
AI VTuber fans' engagement marks not only ``more participation,'' but a qualitative shift from consumption and after-the-fact remix to real-time co-performance. In AI VTuber streams, audience input is instantiated immediately as performance: low-barrier Chat sustains an always-on feedback loop, while SuperChat escalates that loop by pricing the right to steer the unfolding interaction. This intensifies Jenkins' notion of participatory culture \citep{jenkins2008convergence} from community circulation and secondary creation to live, process-oriented co-authoring of the show itself. This mechanism directly explains our RQ3 results (paid prompts as content steering) and reframes RQ1 and RQ2 (attraction/attachment) as the process pleasures of co-performance rather than passive spectatorship.

Our case extends prior work on monetized livestreaming \citep{Lin04072025} that documents visibility hierarchies and stratified participation in human-led streams: whereas paid messages there largely purchase attention and status among peers, in AI VTubing they purchase procedural leverage over the model, i.e., a paid, real-time capacity to condition the model's next action space. Depending on platform data-retention practices, such interventions may even influence future responses. Payment no longer buys a spotlight around preexisting content; it buys a handle on content generation.

Viewed through critical media theory, this empowerment coexists with commodification. What feels like playful prompting is also immaterial labor \citep{terranova2012free}: the production of meaning, affect, and atmosphere that keeps the stream valuable. Following Fuchs' critique of prosumption \citep{fuchs2014digital}, the very agency that enables fans to co-create is immediately stratified and monetized: platforms capture the free labor of open Chat while selling agency via SuperChat. Such monetization no longer buys visibility or recognition; it buys control over AI system behavior in real time. Building on affordances-in-practice \citep{doi:10.1177/1461444818756290}, this configuration shows how a single feature affords different action-capabilities across contexts: in AI VTubing, the affordance of SuperChat is not just ``to be seen'' but ``to make the agent act.''

We term this configuration \emph{real-time co-performance commodification}: platforms convert audience capacity to shape model outputs into a tradable privilege, advancing participatory culture while exposing its commercialization and the broader shift from an attention economy to an engagement economy \citep{nieborg2018platformization, doi:10.1177/2056305119883428}.  In this sense, AI VTubers both intensify participatory culture (extending it from remix to real-time co-performance) and simultaneously expose its inner economic logic, showing that participation is never simply pure empowerment, but is always already entangled with commodification. The design problem that follows is structural rather than cosmetic: future AI-mediated communities should negotiate openness (low-barrier co-performance for all) against revenue (priced intervention for some), making explicit whether, how, and to what extent the right to co-create should be rationed by money. This negotiation is not cosmetic interface design but a structural governance choice about who is entitled to shape AI-mediated performance.

\subsection{Transparent Parasocial Relationships: Consistency as the New Authenticity}
The emergence of AI VTubers profoundly challenges and reconstructs traditional understandings of authenticity in parasocial relationships. For human VTubers, authenticity hinges on a fragile balance between the virtual persona and the Nakanohito \citep{lu2021more}. By contrast,  our findings reveal that AI VTuber fans do not measure authenticity by fidelity to human likeness. Instead, they elevate persona consistency (e.g., stability, coherence, the absence of ``out-of-character'' slips) as a new benchmark of authenticity.

This marks not merely a shift of emphasis but a paradigmatic reversal.  Classic parasocial interaction theory presumes audiences relate to media characters as if they were real humans \citep{schramm2008psi}. AI VTuber fans, however, engage in a \emph{transparent parasocial relationship}: they project emotions and anthropomorphic framings while fully aware of the AI's non-human nature. This reversal is crucial: parasociality here thrives not on illusion but on acknowledged artifice, with consistency and reliability becoming the very grounds of intimacy.

This also extends Enli's theory of mediated authenticity, which frames authenticity as a negotiated achievement between audiences and media figures \citep{enli2015mediated}. In human VTubing, this negotiation is haunted by the instability of performer leakage; in AI VTubing, the absence of Nakanohito becomes a guarantee of stability. Fans perceive persona coherence not as a fragile performance but as structurally secured, and therefore more trustworthy than human realness. Authenticity, in this context, no longer resides in correspondence to lived experience but in the sustained coherence of the persona over time.

Beyond theoretical implications, this reconfiguration provides concrete benchmarks for AI VTuber development. For developers, the task shifts from simulating humanness to engineering systemic reliability: ensuring long-term memory to prevent drift, designing coherent growth trajectories, and building robust guardrails to keep the persona in character. The gold standard shifts from naturalistic imitation to systemic reliability, positioning consistency as the new axis along which both fan attachment and platform trust are built. Crucially, consistency as authenticity is not only a fan expectation but also a platform responsibility, since persona breakdowns risk undermining both emotional bonds and the monetization structures built upon them.

\subsection{Design Implications for Future AI VTubers}
Building on our analysis of co-performance commodification and transparent parasociality, we distill three design imperatives from our findings: (1) monetization strategies that balance revenue with fairness, (2) technical and content design that prioritizes persona consistency while preserving unpredictability, and (3) ethical safeguards that address psychological risks of over-attachment.

\subsubsection{Balancing Revenue and Fairness}
In the context of AI VTubers, sending SuperChats functions as a mechanism for purchasing opportunities to co-create in the livestream, thereby establishing a more interactive monetization model. While this approach demonstrates potential as a sustainable business model, both AI VTuber developers and streaming platforms must carefully balance economic incentives with fairness. A fully pay-to-participate co-creation environment risks discouraging fans with limited financial means from engaging and undermining the communal ethos that sustains participatory culture. To mitigate this, we recommend that AI VTubers, while prioritizing responses to SuperChats, should also preserve interaction opportunities through regular chat messages, ensuring that non-monetary contributions retain a chance of being acknowledged. Furthermore, livestreaming platforms should support such diverse modes of interaction and ensure transparency in monetization mechanisms. This is not merely a matter of interface design but a structural governance choice about how far platforms should monetize co-creation rights. Designing for such balance not only safeguards the spirit of communal co-performance from being fully commodified but also secures a sustainable and legitimate revenue stream for developers. In practice, this means resisting a drift toward fully monetized co-performance, which risks converting all fan agency into capitalized privilege.

\subsubsection{Designing for Consistency, Interactivity, and Transparency} 
Our findings highlight a dual requirement for AI VTuber design: fans are attracted by unpredictability (RQ1) yet remain loyal because of persona consistency and authenticity (RQ2). 
Developers should therefore balance persona consistency with strategic unpredictability, maintaining a stable personality framework while allowing unexpected responses that create entertainment value. This balance reframes unpredictability not as random error but as a curated affordance that drives ongoing engagement. 
Beyond stability, AI VTubers should be designed as interactive entities rather than static performers. Co-creation is not an add-on but the core of their appeal: audience input must be incorporated into system design as a constitutive feature, with the persona evolving in dialogue with its community. 
Finally, rather than concealing the Nakanohito like human VTubers, AI VTubers should embrace technological transparency and construct explicitly AI-centric narratives. Transparency can itself be a narrative resource: the development process, such as feature rollouts or capability updates, can be staged as content, making fans feel they are accompanying and shaping the AI's growth. Such transparency reinforces the new authenticity benchmark of consistency, allowing audiences to trust the persona precisely because its artificiality is openly acknowledged. This repositions technical evolution not as backstage maintenance but as part of the performance itself.

\subsubsection{Ethical Safeguards for Parasocial Dependency}
Our research reveals that fans' anthropomorphization of AI VTubers can lead to deep emotional dependency, potentially creating psychological and social risks. Simple disclosure that ``this is AI'' is insufficient, as fans demonstrate strong tendencies toward affection and attachment regardless of the entity's artificial nature. Such risks are not incidental but intrinsic to transparent parasociality: even when audiences fully recognize the AI's artificiality, consistency and reliability can intensify rather than weaken emotional bonds. 
This necessitates more sophisticated content governance frameworks. Platforms and developers should implement monitoring systems to identify patterns of over-attachment or unhealthy reliance, and develop intervention protocols when necessary. Additionally, built-in mechanisms should promote healthy boundaries, such as limiting rapid-fire interactions, imposing cooldown periods for intensive SuperChats, or offering resources to users showing signs of excessive dependency. Because platforms directly profit from such attachments, they also bear responsibility to mitigate potential harms rather than externalizing them onto users. Balancing authentic engagement with user psychological well-being therefore requires not only technical safeguards but also ongoing collaboration among developers, psychologists, ethicists, and platform operators. Ultimately, ethical safeguards should not be treated as afterthoughts but as integral design parameters, shaping the legitimacy and sustainability of AI-mediated entertainment.

\subsection{Limitations and Future Work}
This study provides preliminary insights into AI VTuber fan communities, yet several limitations constrain the scope of our claims and point to future directions.

First, our analysis centers on a single prominent case, Neuro-sama. This focus enabled analytical depth but limits generalizability across the growing diversity of AI VTubers. Smaller or stylistically different communities may reveal alternative patterns of attraction, attachment, and monetization. Future work should adopt comparative, multi-case approaches to map these variations and test the robustness of our framework.  Similarly, our interaction log analysis concentrated on English-language streams on Twitch. This design choice minimized platform and cultural confounds but excluded other significant contexts, such as Neuro-sama's audiences speaking in other languages. Cross-cultural and cross-platform comparisons would broaden understanding of how AI VTuber fandom manifests across linguistic, cultural, and regulatory environments.

Second, we leveraged generative AI models to assist with transcription and large-scale coding of Chat and SuperChat data. Specifically, \texttt{Gemini-2.5-Pro} was used for transcription (with participant verification), and \texttt{gpt-4.1-mini} and \texttt{Gemini-2.5-Flash} were employed for coding. While prior studies have validated the reliability of LLM-assisted qualitative analysis \citep{wang2024human, xiao2023supporting, chew2023llm}, risks of misinterpretation remain. To mitigate these, we implemented human verification procedures following established validation protocols \citep{hou2024prompt, tai2024examination}. Nonetheless, results are inevitably mediated by the affordances and limitations of current models. We acknowledge these methodological risks while also highlighting AI-assisted analysis as an emerging research practice worthy of further scrutiny in HCI itself.

\section{Conclusion}
This study provides the first comprehensive examination of fan engagement with AI VTubers, revealing how audiences sustain relationships with non-human VTubers through a blend of technical awareness and anthropomorphic projection. We highlight two key dynamics: co-performance, where audience input becomes the core of entertainment rather than a peripheral activity, and transparent parasociality, where authenticity is reconstructed around persona consistency rather than human likeness. These dynamics underpin a novel monetization model in which fans purchase opportunities to shape content, shifting financial support from appreciation to co-creation. By situating these findings within theories of participatory culture, prosumption, and mediated authenticity, we extend scholarship on VTubers and contribute to broader debates on human–AI relationships in digital entertainment. Our analysis further informs the design of future AI VTubers, emphasizing the need to balance revenue and fairness, sustain consistency while embracing unpredictability, and implement safeguards against over-attachment. As AI-mediated performance becomes increasingly prevalent, understanding how audiences negotiate intimacy, authenticity, and agency with artificial personas will be critical for both researchers and practitioners.

\bibliographystyle{ACM-Reference-Format}
\bibliography{sample-sigconf-authordraft}


\begin{thebibliography}{66}


\ifx \showCODEN    \undefined \def \showCODEN     #1{\unskip}     \fi
\ifx \showISBNx    \undefined \def \showISBNx     #1{\unskip}     \fi
\ifx \showISBNxiii \undefined \def \showISBNxiii  #1{\unskip}     \fi
\ifx \showISSN     \undefined \def \showISSN      #1{\unskip}     \fi
\ifx \showLCCN     \undefined \def \showLCCN      #1{\unskip}     \fi
\ifx \shownote     \undefined \def \shownote      #1{#1}          \fi
\ifx \showarticletitle \undefined \def \showarticletitle #1{#1}   \fi
\ifx \showURL      \undefined \def \showURL       {\relax}        \fi
\providecommand\bibfield[2]{#2}
\providecommand\bibinfo[2]{#2}
\providecommand\natexlab[1]{#1}
\providecommand\showeprint[2][]{arXiv:#2}

\bibitem[Amato et~al\mbox{.}(2024)]%
        {amato2024}
\bibfield{author}{\bibinfo{person}{Natale Amato}, \bibinfo{person}{Berardina De~Carolis}, \bibinfo{person}{Francesco de Gioia}, \bibinfo{person}{Corrado Loglisci}, \bibinfo{person}{Giuseppe Palestra}, {and} \bibinfo{person}{Mario~Nicola Venezia}.} \bibinfo{year}{2024}\natexlab{}.
\newblock \showarticletitle{Can an AI-driven VTuber engage people? The KawAIi Case Study}. In \bibinfo{booktitle}{\emph{SOCIALIZE 2024, CEUR Workshop Proceedings}}.
\newblock


\bibitem[Braun and Clarke(2006)]%
        {braun2006using}
\bibfield{author}{\bibinfo{person}{Virginia Braun} {and} \bibinfo{person}{Victoria Clarke}.} \bibinfo{year}{2006}\natexlab{}.
\newblock \showarticletitle{Using thematic analysis in psychology}.
\newblock \bibinfo{journal}{\emph{Qualitative research in psychology}} \bibinfo{volume}{3}, \bibinfo{number}{2} (\bibinfo{year}{2006}), \bibinfo{pages}{77--101}.
\newblock


\bibitem[Bredikhina and Giard(2022)]%
        {bredikhina2022becoming}
\bibfield{author}{\bibinfo{person}{Liudmila Bredikhina} {and} \bibinfo{person}{Agn{\`e}s Giard}.} \bibinfo{year}{2022}\natexlab{}.
\newblock \showarticletitle{Becoming a virtual cutie: digital cross-dressing in Japan}.
\newblock \bibinfo{journal}{\emph{Convergence}} \bibinfo{volume}{28}, \bibinfo{number}{6} (\bibinfo{year}{2022}), \bibinfo{pages}{1643--1661}.
\newblock


\bibitem[Chang et~al\mbox{.}(2025)]%
        {chang2025artificial}
\bibfield{author}{\bibinfo{person}{Yaping Chang}, \bibinfo{person}{Han Wang}, {and} \bibinfo{person}{Zhenjiang Guo}.} \bibinfo{year}{2025}\natexlab{}.
\newblock \showarticletitle{Artificial intelligence in live streaming: How can virtual streamers bring more sales?}
\newblock \bibinfo{journal}{\emph{Journal of Retailing and Consumer Services}}  \bibinfo{volume}{84} (\bibinfo{year}{2025}), \bibinfo{pages}{104247}.
\newblock


\bibitem[Chen et~al\mbox{.}(2024b)]%
        {chen2024avatars}
\bibfield{author}{\bibinfo{person}{Hongquan Chen}, \bibinfo{person}{Bingjia Shao}, \bibinfo{person}{Xuemei Yang}, \bibinfo{person}{Weiyao Kang}, {and} \bibinfo{person}{Wenfang Fan}.} \bibinfo{year}{2024}\natexlab{b}.
\newblock \showarticletitle{Avatars in live streaming commerce: the influence of anthropomorphism on consumers' willingness to accept virtual live streamers}.
\newblock \bibinfo{journal}{\emph{Computers in Human Behavior}}  \bibinfo{volume}{156} (\bibinfo{year}{2024}), \bibinfo{pages}{108216}.
\newblock


\bibitem[Chen et~al\mbox{.}(2024a)]%
        {chen2024digital}
\bibfield{author}{\bibinfo{person}{Xi Chen}, \bibinfo{person}{Siva~Shankar Ramasamy}, {and} \bibinfo{person}{Bibi She}.} \bibinfo{year}{2024}\natexlab{a}.
\newblock \showarticletitle{Digital human technology in the application of live streaming in social media}.
\newblock \bibinfo{journal}{\emph{Radioelectronic and Computer Systems}} \bibinfo{volume}{2024}, \bibinfo{number}{4} (\bibinfo{year}{2024}), \bibinfo{pages}{34--45}.
\newblock


\bibitem[Chew et~al\mbox{.}(2023)]%
        {chew2023llm}
\bibfield{author}{\bibinfo{person}{Robert Chew}, \bibinfo{person}{John Bollenbacher}, \bibinfo{person}{Michael Wenger}, \bibinfo{person}{Jessica Speer}, {and} \bibinfo{person}{Annice Kim}.} \bibinfo{year}{2023}\natexlab{}.
\newblock \bibinfo{title}{LLM-Assisted Content Analysis: Using Large Language Models to Support Deductive Coding}.
\newblock
\showeprint[arxiv]{2306.14924}~[cs.CL]
\urldef\tempurl%
\url{https://arxiv.org/abs/2306.14924}
\showURL{%
\tempurl}


\bibitem[Chinchilla and Kim(2024)]%
        {chinchilla2024vtuber}
\bibfield{author}{\bibinfo{person}{P Chinchilla} {and} \bibinfo{person}{Jihyun Kim}.} \bibinfo{year}{2024}\natexlab{}.
\newblock \showarticletitle{Vtuber for streamers: Exploring the role of social presence in the visual representation of streamers}.
\newblock \bibinfo{journal}{\emph{Communication Studies}} \bibinfo{volume}{75}, \bibinfo{number}{6} (\bibinfo{year}{2024}), \bibinfo{pages}{844--860}.
\newblock


\bibitem[Cohen(1960)]%
        {cohen1960coefficient}
\bibfield{author}{\bibinfo{person}{Jacob Cohen}.} \bibinfo{year}{1960}\natexlab{}.
\newblock \showarticletitle{A coefficient of agreement for nominal scales}.
\newblock \bibinfo{journal}{\emph{Educational and psychological measurement}} \bibinfo{volume}{20}, \bibinfo{number}{1} (\bibinfo{year}{1960}), \bibinfo{pages}{37--46}.
\newblock


\bibitem[Costa(2018)]%
        {doi:10.1177/1461444818756290}
\bibfield{author}{\bibinfo{person}{Elisabetta Costa}.} \bibinfo{year}{2018}\natexlab{}.
\newblock \showarticletitle{Affordances-in-practice: An ethnographic critique of social media logic and context collapse}.
\newblock \bibinfo{journal}{\emph{New Media \& Society}} \bibinfo{volume}{20}, \bibinfo{number}{10} (\bibinfo{year}{2018}), \bibinfo{pages}{3641--3656}.
\newblock
\showeprint{https://doi.org/10.1177/1461444818756290}
\href{https://doi.org/10.1177/1461444818756290}{doi:\nolinkurl{10.1177/1461444818756290}}
\newblock
\shownote{PMID: 30581356}.


\bibitem[Cunningham and Craig(2019)]%
        {doi:10.1177/2056305119883428}
\bibfield{author}{\bibinfo{person}{Stuart Cunningham} {and} \bibinfo{person}{David Craig}.} \bibinfo{year}{2019}\natexlab{}.
\newblock \showarticletitle{Creator Governance in Social Media Entertainment}.
\newblock \bibinfo{journal}{\emph{Social Media + Society}} \bibinfo{volume}{5}, \bibinfo{number}{4} (\bibinfo{year}{2019}), \bibinfo{pages}{2056305119883428}.
\newblock
\showeprint{https://doi.org/10.1177/2056305119883428}
\href{https://doi.org/10.1177/2056305119883428}{doi:\nolinkurl{10.1177/2056305119883428}}


\bibitem[Enli(2015)]%
        {enli2015mediated}
\bibfield{author}{\bibinfo{person}{Gunn Enli}.} \bibinfo{year}{2015}\natexlab{}.
\newblock \bibinfo{booktitle}{\emph{Mediated Authenticity: How the Media Constructs Reality}}.
\newblock
\showISBNx{ISBN 978-1-4331-1485-4}
\href{https://doi.org/10.3726/978-1-4539-1458-8}{doi:\nolinkurl{10.3726/978-1-4539-1458-8}}


\bibitem[Epley et~al\mbox{.}(2007)]%
        {epley2007seeing}
\bibfield{author}{\bibinfo{person}{Nicholas Epley}, \bibinfo{person}{Adam Waytz}, {and} \bibinfo{person}{John~T Cacioppo}.} \bibinfo{year}{2007}\natexlab{}.
\newblock \showarticletitle{On seeing human: a three-factor theory of anthropomorphism.}
\newblock \bibinfo{journal}{\emph{Psychological review}} \bibinfo{volume}{114}, \bibinfo{number}{4} (\bibinfo{year}{2007}), \bibinfo{pages}{864}.
\newblock


\bibitem[Feng et~al\mbox{.}(2022)]%
        {feng2022does}
\bibfield{author}{\bibinfo{person}{Yuanyue Feng}, \bibinfo{person}{Xiaona Li}, {and} \bibinfo{person}{Rongkai Zhang}.} \bibinfo{year}{2022}\natexlab{}.
\newblock \showarticletitle{Does an AI streamer have feelings? The influence of the positive emotions of AI streamer on consumers’ purchase intention}.
\newblock  (\bibinfo{year}{2022}).
\newblock


\bibitem[Fink(2024)]%
        {fink2024conduct}
\bibfield{author}{\bibinfo{person}{Arlene Fink}.} \bibinfo{year}{2024}\natexlab{}.
\newblock \bibinfo{booktitle}{\emph{How to conduct surveys: A step-by-step guide}}.
\newblock \bibinfo{publisher}{SAGE publications}.
\newblock


\bibitem[Fuchs(2014)]%
        {fuchs2014digital}
\bibfield{author}{\bibinfo{person}{Christian Fuchs}.} \bibinfo{year}{2014}\natexlab{}.
\newblock \showarticletitle{Digital prosumption labour on social media in the context of the capitalist regime of time}.
\newblock \bibinfo{journal}{\emph{Time \& Society}} \bibinfo{volume}{23}, \bibinfo{number}{1} (\bibinfo{year}{2014}), \bibinfo{pages}{97--123}.
\newblock


\bibitem[Gao et~al\mbox{.}(2024)]%
        {gao2023build}
\bibfield{author}{\bibinfo{person}{Fengsen Gao}, \bibinfo{person}{Chengjie Dai}, \bibinfo{person}{Ke Fang}, \bibinfo{person}{Yunxuan Li}, \bibinfo{person}{Ji Li}, {and} \bibinfo{person}{Wai Kin~(Victor) Chan}.} \bibinfo{year}{2024}\natexlab{}.
\newblock \showarticletitle{Build Belonging and Trust Proactively: A Humanized Intelligent Streamer Assistant with Personality, Emotion and Memory}. In \bibinfo{booktitle}{\emph{HCI International 2023 -- Late Breaking Posters}}, \bibfield{editor}{\bibinfo{person}{Constantine Stephanidis}, \bibinfo{person}{Margherita Antona}, \bibinfo{person}{Stavroula Ntoa}, {and} \bibinfo{person}{Gavriel Salvendy}} (Eds.). \bibinfo{publisher}{Springer Nature Switzerland}, \bibinfo{address}{Cham}, \bibinfo{pages}{140--147}.
\newblock
\showISBNx{978-3-031-49215-0}


\bibitem[Hogg(2016)]%
        {hogg2016social}
\bibfield{author}{\bibinfo{person}{Michael~A. Hogg}.} \bibinfo{year}{2016}\natexlab{}.
\newblock \bibinfo{booktitle}{\emph{Social Identity Theory}}.
\newblock \bibinfo{publisher}{Springer International Publishing}, \bibinfo{address}{Cham}, \bibinfo{pages}{3--17}.
\newblock
\showISBNx{978-3-319-29869-6}
\href{https://doi.org/10.1007/978-3-319-29869-6_1}{doi:\nolinkurl{10.1007/978-3-319-29869-6_1}}


\bibitem[Hou et~al\mbox{.}(2024)]%
        {hou2024prompt}
\bibfield{author}{\bibinfo{person}{Chenyu Hou}, \bibinfo{person}{Gaoxia Zhu}, \bibinfo{person}{Juan Zheng}, \bibinfo{person}{Lishan Zhang}, \bibinfo{person}{Xiaoshan Huang}, \bibinfo{person}{Tianlong Zhong}, \bibinfo{person}{Shan Li}, \bibinfo{person}{Hanxiang Du}, {and} \bibinfo{person}{Chin~Lee Ker}.} \bibinfo{year}{2024}\natexlab{}.
\newblock \showarticletitle{Prompt-based and Fine-tuned GPT Models for Context-Dependent and -Independent Deductive Coding in Social Annotation}. In \bibinfo{booktitle}{\emph{Proceedings of the 14th Learning Analytics and Knowledge Conference}} (Kyoto, Japan) \emph{(\bibinfo{series}{LAK '24})}. \bibinfo{publisher}{Association for Computing Machinery}, \bibinfo{address}{New York, NY, USA}, \bibinfo{pages}{518–528}.
\newblock
\showISBNx{9798400716188}
\href{https://doi.org/10.1145/3636555.3636910}{doi:\nolinkurl{10.1145/3636555.3636910}}


\bibitem[Houssard et~al\mbox{.}(2023)]%
        {houssard2023monetization}
\bibfield{author}{\bibinfo{person}{Antoine Houssard}, \bibinfo{person}{Federico Pilati}, \bibinfo{person}{Maria Tartari}, \bibinfo{person}{Pier~Luigi Sacco}, {and} \bibinfo{person}{Riccardo Gallotti}.} \bibinfo{year}{2023}\natexlab{}.
\newblock \showarticletitle{Monetization in online streaming platforms: an exploration of inequalities in Twitch. tv}.
\newblock \bibinfo{journal}{\emph{Scientific Reports}} \bibinfo{volume}{13}, \bibinfo{number}{1} (\bibinfo{year}{2023}), \bibinfo{pages}{1103}.
\newblock


\bibitem[Hu and Ma(2023)]%
        {hu2023human}
\bibfield{author}{\bibinfo{person}{Hai-hua Hu} {and} \bibinfo{person}{Fang Ma}.} \bibinfo{year}{2023}\natexlab{}.
\newblock \showarticletitle{Human-like bots are not humans: The weakness of sensory language for virtual streamers in livestream commerce}.
\newblock \bibinfo{journal}{\emph{Journal of Retailing and Consumer Services}}  \bibinfo{volume}{75} (\bibinfo{year}{2023}), \bibinfo{pages}{103541}.
\newblock


\bibitem[Jenkins and Deuze(2008)]%
        {jenkins2008convergence}
\bibfield{author}{\bibinfo{person}{Henry Jenkins} {and} \bibinfo{person}{Mark Deuze}.} \bibinfo{year}{2008}\natexlab{}.
\newblock \bibinfo{title}{Convergence culture}.
\newblock \bibinfo{numpages}{5--12}~pages.
\newblock


\bibitem[Jiang et~al\mbox{.}(2025)]%
        {jiang2025smile}
\bibfield{author}{\bibinfo{person}{Kan Jiang}, \bibinfo{person}{Meilian Qin}, \bibinfo{person}{Dejun Deng}, {and} \bibinfo{person}{Dailan Zhou}.} \bibinfo{year}{2025}\natexlab{}.
\newblock \showarticletitle{Smile or Not Smile: The Effect of Virtual Influencers' Emotional Expression on Brand Authenticity, Purchase Intention and Follow Intention}.
\newblock \bibinfo{journal}{\emph{Journal of Consumer Behaviour}} \bibinfo{volume}{24}, \bibinfo{number}{2} (\bibinfo{year}{2025}), \bibinfo{pages}{962--981}.
\newblock


\bibitem[Kim et~al\mbox{.}(2025)]%
        {kim2025vtuber}
\bibfield{author}{\bibinfo{person}{Daye Kim}, \bibinfo{person}{Sebin Lee}, \bibinfo{person}{Yoonseo Jun}, \bibinfo{person}{Yujin Shin}, {and} \bibinfo{person}{Jungjin Lee}.} \bibinfo{year}{2025}\natexlab{}.
\newblock \showarticletitle{VTuber's Atelier: The Design Space, Challenges, and Opportunities for VTubing}. In \bibinfo{booktitle}{\emph{Proceedings of the 2025 CHI Conference on Human Factors in Computing Systems}} \emph{(\bibinfo{series}{CHI '25})}. \bibinfo{publisher}{Association for Computing Machinery}, \bibinfo{address}{New York, NY, USA}, Article \bibinfo{articleno}{1242}, \bibinfo{numpages}{23}~pages.
\newblock
\showISBNx{9798400713941}
\href{https://doi.org/10.1145/3706598.3714107}{doi:\nolinkurl{10.1145/3706598.3714107}}


\bibitem[{Kizuna AI}(2016)]%
        {kizunaai}
\bibfield{author}{\bibinfo{person}{{Kizuna AI}}.} \bibinfo{year}{2016}\natexlab{}.
\newblock \bibinfo{booktitle}{\emph{A.I.Channel}}.
\newblock YouTube.
\newblock
\urldef\tempurl%
\url{https://www.youtube.com/channel/UC4YaOt1yT-ZeyB0OmxHgolA}
\showURL{%
\tempurl}
\newblock
\shownote{Virtual YouTuber channel launched in 2016}.


\bibitem[Landis and Koch(1977)]%
        {landis1977measurement}
\bibfield{author}{\bibinfo{person}{J~Richard Landis} {and} \bibinfo{person}{Gary~G Koch}.} \bibinfo{year}{1977}\natexlab{}.
\newblock \showarticletitle{The measurement of observer agreement for categorical data}.
\newblock \bibinfo{journal}{\emph{biometrics}} (\bibinfo{year}{1977}), \bibinfo{pages}{159--174}.
\newblock


\bibitem[Lee et~al\mbox{.}(2025)]%
        {lee2025can}
\bibfield{author}{\bibinfo{person}{Ken~Jen Lee}, \bibinfo{person}{PiaoHong Wang}, {and} \bibinfo{person}{Zhicong Lu}.} \bibinfo{year}{2025}\natexlab{}.
\newblock \showarticletitle{"Can't believe I'm crying over an anime girl": Public Parasocial Grieving and Coping Towards VTuber Graduation and Termination}. In \bibinfo{booktitle}{\emph{Proceedings of the 2025 CHI Conference on Human Factors in Computing Systems}} \emph{(\bibinfo{series}{CHI '25})}. \bibinfo{publisher}{Association for Computing Machinery}, \bibinfo{address}{New York, NY, USA}, Article \bibinfo{articleno}{1237}, \bibinfo{numpages}{23}~pages.
\newblock
\showISBNx{9798400713941}
\href{https://doi.org/10.1145/3706598.3714216}{doi:\nolinkurl{10.1145/3706598.3714216}}


\bibitem[Lee and Lee(2023)]%
        {lee2023ju}
\bibfield{author}{\bibinfo{person}{Sebin Lee} {and} \bibinfo{person}{Jungjin Lee}.} \bibinfo{year}{2023}\natexlab{}.
\newblock \showarticletitle{“Ju. T’aime” my idol, my streamer: A case study on fandom experience as audiences and creators of VTuber concert}.
\newblock \bibinfo{journal}{\emph{IEEE Access}}  \bibinfo{volume}{11} (\bibinfo{year}{2023}), \bibinfo{pages}{31125--31142}.
\newblock


\bibitem[Lerman and Yitzhaki(1984)]%
        {lerman1984note}
\bibfield{author}{\bibinfo{person}{Robert~I Lerman} {and} \bibinfo{person}{Shlomo Yitzhaki}.} \bibinfo{year}{1984}\natexlab{}.
\newblock \showarticletitle{A note on the calculation and interpretation of the Gini index}.
\newblock \bibinfo{journal}{\emph{Economics Letters}} \bibinfo{volume}{15}, \bibinfo{number}{3-4} (\bibinfo{year}{1984}), \bibinfo{pages}{363--368}.
\newblock


\bibitem[Li(2023)]%
        {li2023does}
\bibfield{author}{\bibinfo{person}{Yijin Li}.} \bibinfo{year}{2023}\natexlab{}.
\newblock \showarticletitle{Why does Gen Z watch virtual streaming VTube anime videos with avatars on Twitch?}
\newblock \bibinfo{journal}{\emph{Online Media and Global Communication}} \bibinfo{volume}{2}, \bibinfo{number}{3} (\bibinfo{year}{2023}), \bibinfo{pages}{379--403}.
\newblock


\bibitem[Li and Guo(2021)]%
        {li2021virtual}
\bibfield{author}{\bibinfo{person}{Yi Li} {and} \bibinfo{person}{Yunjun Guo}.} \bibinfo{year}{2021}\natexlab{}.
\newblock \showarticletitle{Virtual gifting and danmaku: What motivates people to interact in game live streaming?}
\newblock \bibinfo{journal}{\emph{Telematics and Informatics}}  \bibinfo{volume}{62} (\bibinfo{year}{2021}), \bibinfo{pages}{101624}.
\newblock


\bibitem[Li et~al\mbox{.}(2023)]%
        {li2023blibug}
\bibfield{author}{\bibinfo{person}{Yihua Li}, \bibinfo{person}{Yuqian Sun}, \bibinfo{person}{Ying Xu}, {and} \bibinfo{person}{Jihong Yu}.} \bibinfo{year}{2023}\natexlab{}.
\newblock \showarticletitle{Blibug: AI Vtuber Based on Bilibili Danmuku Interaction}. In \bibinfo{booktitle}{\emph{Proceedings of the 15th Conference on Creativity and Cognition}}. \bibinfo{pages}{387--390}.
\newblock


\bibitem[Lin(2025)]%
        {Lin04072025}
\bibfield{author}{\bibinfo{person}{Hui Lin}.} \bibinfo{year}{2025}\natexlab{}.
\newblock \showarticletitle{‘Let’s purchase coloured live chat messages’: the impact of user engagement with Super Chat on YouTube}.
\newblock \bibinfo{journal}{\emph{Information, Communication \& Society}} \bibinfo{volume}{28}, \bibinfo{number}{9} (\bibinfo{year}{2025}), \bibinfo{pages}{1608--1626}.
\newblock
\showeprint{https://doi.org/10.1080/1369118X.2024.2442407}
\href{https://doi.org/10.1080/1369118X.2024.2442407}{doi:\nolinkurl{10.1080/1369118X.2024.2442407}}


\bibitem[Liu et~al\mbox{.}(2025)]%
        {liu2025impact}
\bibfield{author}{\bibinfo{person}{Hao Liu}, \bibinfo{person}{Peilin Zhang}, \bibinfo{person}{Hongqing Cheng}, \bibinfo{person}{Najmul Hasan}, {and} \bibinfo{person}{Raymond Chiong}.} \bibinfo{year}{2025}\natexlab{}.
\newblock \showarticletitle{Impact of AI-generated virtual streamer interaction on consumer purchase intention: A focus on social presence and perceived value}.
\newblock \bibinfo{journal}{\emph{Journal of Retailing and Consumer Services}}  \bibinfo{volume}{85} (\bibinfo{year}{2025}), \bibinfo{pages}{104290}.
\newblock


\bibitem[Lu et~al\mbox{.}(2021)]%
        {lu2021more}
\bibfield{author}{\bibinfo{person}{Zhicong Lu}, \bibinfo{person}{Chenxinran Shen}, \bibinfo{person}{Jiannan Li}, \bibinfo{person}{Hong Shen}, {and} \bibinfo{person}{Daniel Wigdor}.} \bibinfo{year}{2021}\natexlab{}.
\newblock \showarticletitle{More Kawaii than a Real-Person Live Streamer: Understanding How the Otaku Community Engages with and Perceives Virtual YouTubers}. In \bibinfo{booktitle}{\emph{Proceedings of the 2021 CHI Conference on Human Factors in Computing Systems}} (Yokohama, Japan) \emph{(\bibinfo{series}{CHI '21})}. \bibinfo{publisher}{Association for Computing Machinery}, \bibinfo{address}{New York, NY, USA}, Article \bibinfo{articleno}{137}, \bibinfo{numpages}{14}~pages.
\newblock
\showISBNx{9781450380966}
\href{https://doi.org/10.1145/3411764.3445660}{doi:\nolinkurl{10.1145/3411764.3445660}}


\bibitem[Lu et~al\mbox{.}(2018)]%
        {lu2018you}
\bibfield{author}{\bibinfo{person}{Zhicong Lu}, \bibinfo{person}{Haijun Xia}, \bibinfo{person}{Seongkook Heo}, {and} \bibinfo{person}{Daniel Wigdor}.} \bibinfo{year}{2018}\natexlab{}.
\newblock \showarticletitle{You Watch, You Give, and You Engage: A Study of Live Streaming Practices in China}. In \bibinfo{booktitle}{\emph{Proceedings of the 2018 CHI Conference on Human Factors in Computing Systems}} (Montreal QC, Canada) \emph{(\bibinfo{series}{CHI '18})}. \bibinfo{publisher}{Association for Computing Machinery}, \bibinfo{address}{New York, NY, USA}, \bibinfo{pages}{1–13}.
\newblock
\showISBNx{9781450356206}
\href{https://doi.org/10.1145/3173574.3174040}{doi:\nolinkurl{10.1145/3173574.3174040}}


\bibitem[Nieborg and Poell(2018)]%
        {nieborg2018platformization}
\bibfield{author}{\bibinfo{person}{David~B Nieborg} {and} \bibinfo{person}{Thomas Poell}.} \bibinfo{year}{2018}\natexlab{}.
\newblock \showarticletitle{The platformization of cultural production: Theorizing the contingent cultural commodity}.
\newblock \bibinfo{journal}{\emph{New media \& society}} \bibinfo{volume}{20}, \bibinfo{number}{11} (\bibinfo{year}{2018}), \bibinfo{pages}{4275--4292}.
\newblock


\bibitem[Peng et~al\mbox{.}(2024)]%
        {peng2024impact}
\bibfield{author}{\bibinfo{person}{Yuhong Peng}, \bibinfo{person}{Yedi Wang}, \bibinfo{person}{Jingpeng Li}, {and} \bibinfo{person}{Qiang Yang}.} \bibinfo{year}{2024}\natexlab{}.
\newblock \showarticletitle{Impact of AI-oriented live-streaming E-commerce service failures on consumer disengagement—empirical evidence from China}.
\newblock \bibinfo{journal}{\emph{Journal of Theoretical and Applied Electronic Commerce Research}} \bibinfo{volume}{19}, \bibinfo{number}{2} (\bibinfo{year}{2024}), \bibinfo{pages}{1580--1598}.
\newblock


\bibitem[Peres et~al\mbox{.}(2024)]%
        {PERES2024403}
\bibfield{author}{\bibinfo{person}{Renana Peres}, \bibinfo{person}{Martin Schreier}, \bibinfo{person}{David~A. Schweidel}, {and} \bibinfo{person}{Alina Sorescu}.} \bibinfo{year}{2024}\natexlab{}.
\newblock \showarticletitle{The creator economy: An introduction and a call for scholarly research}.
\newblock \bibinfo{journal}{\emph{International Journal of Research in Marketing}} \bibinfo{volume}{41}, \bibinfo{number}{3} (\bibinfo{year}{2024}), \bibinfo{pages}{403--410}.
\newblock
\showISSN{0167-8116}
\href{https://doi.org/10.1016/j.ijresmar.2024.07.005}{doi:\nolinkurl{10.1016/j.ijresmar.2024.07.005}}


\bibitem[Renella(2023)]%
        {renella2023machine}
\bibfield{author}{\bibinfo{person}{Noah Renella}.} \bibinfo{year}{2023}\natexlab{}.
\newblock \showarticletitle{Machine learning models for assisting twitch streamers}.
\newblock \bibinfo{journal}{\emph{URL: https://scholarworks. calstate. edu/downloads/vx021n95n}} (\bibinfo{year}{2023}).
\newblock


\bibitem[Ribeiro et~al\mbox{.}(2024)]%
        {ribeiro2024towards}
\bibfield{author}{\bibinfo{person}{Ailton Ribeiro}, \bibinfo{person}{Murilo Arouca}, \bibinfo{person}{Ana Amorim}, \bibinfo{person}{Maria Pestana}, {and} \bibinfo{person}{Vaninha Vieira}.} \bibinfo{year}{2024}\natexlab{}.
\newblock \showarticletitle{Towards Inclusive Avatars: A Study on Self-Representation in Virtual Environments}. In \bibinfo{booktitle}{\emph{Anais do XIX Simpósio Brasileiro de Sistemas Colaborativos}} (Salvador/BA). \bibinfo{publisher}{SBC}, \bibinfo{address}{Porto Alegre, RS, Brasil}, \bibinfo{pages}{13--27}.
\newblock
\showISSN{2326-2842}
\href{https://doi.org/10.5753/sbsc.2024.238056}{doi:\nolinkurl{10.5753/sbsc.2024.238056}}


\bibitem[Rohrbacher and Mishra(2024)]%
        {rohrbacher2024vtubing}
\bibfield{author}{\bibinfo{person}{Patricia Rohrbacher} {and} \bibinfo{person}{Deepti Mishra}.} \bibinfo{year}{2024}\natexlab{}.
\newblock \showarticletitle{VTubing and Its Potential for the Streaming and Design Community: An Austrian Perspective}. In \bibinfo{booktitle}{\emph{Social Computing and Social Media}}, \bibfield{editor}{\bibinfo{person}{Adela Coman} {and} \bibinfo{person}{Simona Vasilache}} (Eds.). \bibinfo{publisher}{Springer Nature Switzerland}, \bibinfo{address}{Cham}, \bibinfo{pages}{222--233}.
\newblock
\showISBNx{978-3-031-61281-7}


\bibitem[Salles et~al\mbox{.}(2020)]%
        {salles2020anthropomorphism}
\bibfield{author}{\bibinfo{person}{Arleen Salles}, \bibinfo{person}{Kathinka Evers}, {and} \bibinfo{person}{Michele Farisco}.} \bibinfo{year}{2020}\natexlab{}.
\newblock \showarticletitle{Anthropomorphism in AI}.
\newblock \bibinfo{journal}{\emph{AJOB neuroscience}} \bibinfo{volume}{11}, \bibinfo{number}{2} (\bibinfo{year}{2020}), \bibinfo{pages}{88--95}.
\newblock


\bibitem[Sanyoura and Anderson(2022)]%
        {el2022quantifying}
\bibfield{author}{\bibinfo{person}{Lana~El Sanyoura} {and} \bibinfo{person}{Ashton Anderson}.} \bibinfo{year}{2022}\natexlab{}.
\newblock \showarticletitle{Quantifying the Creator Economy: A Large-Scale Analysis of Patreon}.
\newblock \bibinfo{journal}{\emph{Proceedings of the International AAAI Conference on Web and Social Media}} \bibinfo{volume}{16}, \bibinfo{number}{1} (\bibinfo{date}{May} \bibinfo{year}{2022}), \bibinfo{pages}{829--840}.
\newblock
\href{https://doi.org/10.1609/icwsm.v16i1.19338}{doi:\nolinkurl{10.1609/icwsm.v16i1.19338}}


\bibitem[Schramm and Hartmann(2008)]%
        {schramm2008psi}
\bibfield{author}{\bibinfo{person}{Holger Schramm} {and} \bibinfo{person}{Tilo Hartmann}.} \bibinfo{year}{2008}\natexlab{}.
\newblock \showarticletitle{The PSI-Process Scales. A new measure to assess the intensity and breadth of parasocial processes}.
\newblock  (\bibinfo{year}{2008}).
\newblock


\bibitem[Stein et~al\mbox{.}(2024)]%
        {stein2024parasocial}
\bibfield{author}{\bibinfo{person}{Jan-Philipp Stein}, \bibinfo{person}{Priska Linda~Breves}, {and} \bibinfo{person}{Nora Anders}.} \bibinfo{year}{2024}\natexlab{}.
\newblock \showarticletitle{Parasocial interactions with real and virtual influencers: The role of perceived similarity and human-likeness}.
\newblock \bibinfo{journal}{\emph{New Media \& Society}} \bibinfo{volume}{26}, \bibinfo{number}{6} (\bibinfo{year}{2024}), \bibinfo{pages}{3433--3453}.
\newblock


\bibitem[Sutandijo and Qomariyah(2023)]%
        {sutandijo2023artificial}
\bibfield{author}{\bibinfo{person}{Vincent~Joyan Sutandijo} {and} \bibinfo{person}{Nunung~Nurul Qomariyah}.} \bibinfo{year}{2023}\natexlab{}.
\newblock \showarticletitle{Artificial intelligence based automatic live stream chat machine translator}.
\newblock \bibinfo{journal}{\emph{Procedia Computer Science}}  \bibinfo{volume}{227} (\bibinfo{year}{2023}), \bibinfo{pages}{454--463}.
\newblock


\bibitem[Tai et~al\mbox{.}(2024)]%
        {tai2024examination}
\bibfield{author}{\bibinfo{person}{Robert~H Tai}, \bibinfo{person}{Lillian~R Bentley}, \bibinfo{person}{Xin Xia}, \bibinfo{person}{Jason~M Sitt}, \bibinfo{person}{Sarah~C Fankhauser}, \bibinfo{person}{Ana~M Chicas-Mosier}, {and} \bibinfo{person}{Barnas~G Monteith}.} \bibinfo{year}{2024}\natexlab{}.
\newblock \showarticletitle{An examination of the use of large language models to aid analysis of textual data}.
\newblock \bibinfo{journal}{\emph{International Journal of Qualitative Methods}}  \bibinfo{volume}{23} (\bibinfo{year}{2024}), \bibinfo{pages}{16094069241231168}.
\newblock


\bibitem[Tavakol and Dennick(2011)]%
        {tavakol2011making}
\bibfield{author}{\bibinfo{person}{Mohsen Tavakol} {and} \bibinfo{person}{Reg Dennick}.} \bibinfo{year}{2011}\natexlab{}.
\newblock \showarticletitle{Making sense of Cronbach's alpha}.
\newblock \bibinfo{journal}{\emph{International journal of medical education}}  \bibinfo{volume}{2} (\bibinfo{year}{2011}), \bibinfo{pages}{53}.
\newblock


\bibitem[Terranova(2012)]%
        {terranova2012free}
\bibfield{author}{\bibinfo{person}{Tiziana Terranova}.} \bibinfo{year}{2012}\natexlab{}.
\newblock \showarticletitle{Free labor}.
\newblock In \bibinfo{booktitle}{\emph{Digital labor}}. \bibinfo{publisher}{Routledge}, \bibinfo{pages}{33--57}.
\newblock


\bibitem[{User Local}(2022)]%
        {userlocal_press_2022}
\bibfield{author}{\bibinfo{person}{{User Local}}.} \bibinfo{year}{2022}\natexlab{}.
\newblock \bibinfo{booktitle}{\emph{Virtual Talent Popularity Ranking "VTuber Database" Celebrates its 4th Anniversary}}.
\newblock
\urldef\tempurl%
\url{https://www.userlocal.jp/press/20221129vt/}
\showURL{%
\tempurl}


\bibitem[{User Local}(2025)]%
        {userlocal_vtuber_ranking}
\bibfield{author}{\bibinfo{person}{{User Local}}.} \bibinfo{year}{2025}\natexlab{}.
\newblock \bibinfo{booktitle}{\emph{VTuber Ranking}}.
\newblock
\urldef\tempurl%
\url{https://virtual-youtuber.userlocal.jp/document/ranking}
\showURL{%
\tempurl}


\bibitem[{VTuber Awards}(2024)]%
        {vtuberawards2024}
\bibfield{author}{\bibinfo{person}{{VTuber Awards}}.} \bibinfo{year}{2024}\natexlab{}.
\newblock \bibinfo{booktitle}{\emph{The {VTuber} {Awards} 2024 Winners}}.
\newblock The VTuber Awards.
\newblock
\urldef\tempurl%
\url{https://www.thevtuberawards.com/winners/2024}
\showURL{%
\tempurl}
\newblock
\shownote{Neuro-sama recognized among prominent VTubers in the 2024 VTuber Awards}.


\bibitem[Wan and Lu(2024)]%
        {wan2024investigating}
\bibfield{author}{\bibinfo{person}{Qian Wan} {and} \bibinfo{person}{Zhicong Lu}.} \bibinfo{year}{2024}\natexlab{}.
\newblock \showarticletitle{Investigating vtubing as a reconstruction of streamer self-presentation: Identity, performance, and gender}.
\newblock \bibinfo{journal}{\emph{Proceedings of the ACM on human-computer interaction}} \bibinfo{volume}{8}, \bibinfo{number}{CSCW1} (\bibinfo{year}{2024}), \bibinfo{pages}{1--22}.
\newblock


\bibitem[Wang et~al\mbox{.}(2023)]%
        {wang2023role}
\bibfield{author}{\bibinfo{person}{Lingli Wang}, \bibinfo{person}{Yumei He}, \bibinfo{person}{Ni Huang}, \bibinfo{person}{De Liu}, \bibinfo{person}{Xunhua Guo}, {and} \bibinfo{person}{Guoqing Chen}.} \bibinfo{year}{2023}\natexlab{}.
\newblock \showarticletitle{The role of AI assistants in livestream selling: Evidence from a randomized field experiment}.
\newblock \bibinfo{journal}{\emph{University of Miami Business School Research Paper}} \bibinfo{number}{4365103} (\bibinfo{year}{2023}).
\newblock


\bibitem[Wang et~al\mbox{.}(2024)]%
        {wang2024human}
\bibfield{author}{\bibinfo{person}{Xinru Wang}, \bibinfo{person}{Hannah Kim}, \bibinfo{person}{Sajjadur Rahman}, \bibinfo{person}{Kushan Mitra}, {and} \bibinfo{person}{Zhengjie Miao}.} \bibinfo{year}{2024}\natexlab{}.
\newblock \showarticletitle{Human-LLM Collaborative Annotation Through Effective Verification of LLM Labels}. In \bibinfo{booktitle}{\emph{Proceedings of the 2024 CHI Conference on Human Factors in Computing Systems}} (Honolulu, HI, USA) \emph{(\bibinfo{series}{CHI '24})}. \bibinfo{publisher}{Association for Computing Machinery}, \bibinfo{address}{New York, NY, USA}, Article \bibinfo{articleno}{303}, \bibinfo{numpages}{21}~pages.
\newblock
\showISBNx{9798400703300}
\href{https://doi.org/10.1145/3613904.3641960}{doi:\nolinkurl{10.1145/3613904.3641960}}


\bibitem[Wei and Tyson(2025)]%
        {wei2025virtual}
\bibfield{author}{\bibinfo{person}{Yiluo Wei} {and} \bibinfo{person}{Gareth Tyson}.} \bibinfo{year}{2025}\natexlab{}.
\newblock \showarticletitle{Virtual Stars, Real Fans: Understanding the VTuber Ecosystem}. In \bibinfo{booktitle}{\emph{Proceedings of the ACM on Web Conference 2025}} (Sydney NSW, Australia) \emph{(\bibinfo{series}{WWW '25})}. \bibinfo{publisher}{Association for Computing Machinery}, \bibinfo{address}{New York, NY, USA}, \bibinfo{pages}{2352–2365}.
\newblock
\showISBNx{9798400712746}
\href{https://doi.org/10.1145/3696410.3714803}{doi:\nolinkurl{10.1145/3696410.3714803}}


\bibitem[Xiao et~al\mbox{.}(2023)]%
        {xiao2023supporting}
\bibfield{author}{\bibinfo{person}{Ziang Xiao}, \bibinfo{person}{Xingdi Yuan}, \bibinfo{person}{Q.~Vera Liao}, \bibinfo{person}{Rania Abdelghani}, {and} \bibinfo{person}{Pierre-Yves Oudeyer}.} \bibinfo{year}{2023}\natexlab{}.
\newblock \showarticletitle{Supporting Qualitative Analysis with Large Language Models: Combining Codebook with GPT-3 for Deductive Coding}. In \bibinfo{booktitle}{\emph{Companion Proceedings of the 28th International Conference on Intelligent User Interfaces}} (Sydney, NSW, Australia) \emph{(\bibinfo{series}{IUI '23 Companion})}. \bibinfo{publisher}{Association for Computing Machinery}, \bibinfo{address}{New York, NY, USA}, \bibinfo{pages}{75–78}.
\newblock
\showISBNx{9798400701078}
\href{https://doi.org/10.1145/3581754.3584136}{doi:\nolinkurl{10.1145/3581754.3584136}}


\bibitem[Xu et~al\mbox{.}(2025)]%
        {xu2025future}
\bibfield{author}{\bibinfo{person}{Bin Xu}, \bibinfo{person}{Omkar Dastane}, \bibinfo{person}{Eugene Cheng-Xi Aw}, {and} \bibinfo{person}{Suchita Jha}.} \bibinfo{year}{2025}\natexlab{}.
\newblock \showarticletitle{The future of live-streaming commerce: understanding the role of AI-powered virtual streamers}.
\newblock \bibinfo{journal}{\emph{Asia Pacific Journal of Marketing and Logistics}} \bibinfo{volume}{37}, \bibinfo{number}{5} (\bibinfo{year}{2025}), \bibinfo{pages}{1175--1196}.
\newblock


\bibitem[Xu(2021)]%
        {xu2021research}
\bibfield{author}{\bibinfo{person}{Si-han Xu}.} \bibinfo{year}{2021}\natexlab{}.
\newblock \showarticletitle{The Research on Applying Artificial Intelligence Technology to Virtual YouTuber}. In \bibinfo{booktitle}{\emph{2021 IEEE International Conference on Robotics, Automation and Artificial Intelligence (RAAI)}}. \bibinfo{pages}{10--14}.
\newblock
\href{https://doi.org/10.1109/RAAI52226.2021.9507778}{doi:\nolinkurl{10.1109/RAAI52226.2021.9507778}}


\bibitem[Yan et~al\mbox{.}(2025)]%
        {yan2025can}
\bibfield{author}{\bibinfo{person}{Rui Yan}, \bibinfo{person}{Zhen Tang}, {and} \bibinfo{person}{Dewen Liu}.} \bibinfo{year}{2025}\natexlab{}.
\newblock \showarticletitle{Can virtual streamers replace human streamers? The interactive effect of streamer type and product type on purchase intention}.
\newblock \bibinfo{journal}{\emph{Marketing Intelligence \& Planning}} \bibinfo{volume}{43}, \bibinfo{number}{2} (\bibinfo{year}{2025}), \bibinfo{pages}{297--322}.
\newblock


\bibitem[Yuan et~al\mbox{.}(2025)]%
        {yuan2025machines}
\bibfield{author}{\bibinfo{person}{Haixia Yuan}, \bibinfo{person}{Kevin L{\"u}}, {and} \bibinfo{person}{Wenting Fang}.} \bibinfo{year}{2025}\natexlab{}.
\newblock \showarticletitle{Machines vs. humans: The evolving role of artificial intelligence in livestreaming e-commerce}.
\newblock \bibinfo{journal}{\emph{Journal of Business Research}}  \bibinfo{volume}{188} (\bibinfo{year}{2025}), \bibinfo{pages}{115077}.
\newblock


\bibitem[Zhan and Zhang(2023)]%
        {zhan2023exploring}
\bibfield{author}{\bibinfo{person}{Jinming Zhan} {and} \bibinfo{person}{Nan Zhang}.} \bibinfo{year}{2023}\natexlab{}.
\newblock \showarticletitle{Exploring the Impact of Virtual Anchor Features and Live Content on Viewers' Willingness to Pay for “Superchat” in Live Entertainment Scenarios}.
\newblock \bibinfo{journal}{\emph{Highlights in Business, Economics and Management}}  \bibinfo{volume}{6} (\bibinfo{year}{2023}), \bibinfo{pages}{189--205}.
\newblock


\bibitem[Zhang et~al\mbox{.}(2024)]%
        {zhang2024virtual}
\bibfield{author}{\bibinfo{person}{Xianfeng Zhang}, \bibinfo{person}{Yuxue Shi}, \bibinfo{person}{Ting Li}, \bibinfo{person}{Yuxian Guan}, {and} \bibinfo{person}{Xinlei Cui}.} \bibinfo{year}{2024}\natexlab{}.
\newblock \showarticletitle{How do virtual AI streamers influence viewers’ livestream shopping behavior? The effects of persuasive factors and the mediating role of arousal}.
\newblock \bibinfo{journal}{\emph{Information Systems Frontiers}} \bibinfo{volume}{26}, \bibinfo{number}{5} (\bibinfo{year}{2024}), \bibinfo{pages}{1803--1834}.
\newblock


\bibitem[Zhao et~al\mbox{.}(2025)]%
        {zhao2025reaps}
\bibfield{author}{\bibinfo{person}{Ruijing Zhao}, \bibinfo{person}{Brian Diep}, \bibinfo{person}{Jiaxin Pei}, \bibinfo{person}{Dongwook Yoon}, \bibinfo{person}{David Jurgens}, {and} \bibinfo{person}{Jian Zhu}.} \bibinfo{year}{2025}\natexlab{}.
\newblock \showarticletitle{Who Reaps All the Superchats? A Large-Scale Analysis of Income Inequality in Virtual YouTuber Livestreaming}. In \bibinfo{booktitle}{\emph{Proceedings of the 2025 CHI Conference on Human Factors in Computing Systems}} \emph{(\bibinfo{series}{CHI '25})}. \bibinfo{publisher}{Association for Computing Machinery}, \bibinfo{address}{New York, NY, USA}, Article \bibinfo{articleno}{1056}, \bibinfo{numpages}{18}~pages.
\newblock
\showISBNx{9798400713941}
\href{https://doi.org/10.1145/3706598.3713877}{doi:\nolinkurl{10.1145/3706598.3713877}}


\bibitem[Zhu et~al\mbox{.}(2025)]%
        {zhu2025effects}
\bibfield{author}{\bibinfo{person}{Yu-Peng Zhu}, \bibinfo{person}{Lina Xin}, \bibinfo{person}{Huimin Wang}, {and} \bibinfo{person}{Han-Woo Park}.} \bibinfo{year}{2025}\natexlab{}.
\newblock \showarticletitle{Effects of AI virtual anchors on brand image and loyalty: Insights from perceived value theory and SEM-ANN analysis}.
\newblock \bibinfo{journal}{\emph{Systems}} \bibinfo{volume}{13}, \bibinfo{number}{2} (\bibinfo{year}{2025}), \bibinfo{pages}{79}.
\newblock


\end{thebibliography}

\appendix

\section{Survey Content}
\label{appendix:survey_content}
In this section, we present the full content of our survey.
\begin{enumerate}[label={}]
\item \textbf{I have read the information above and I consent to participate in this survey. }
    \begin{itemize}
        \item Yes
        \item No
    \end{itemize}

\item \textbf{1.1 Do you watch VTubers (Virtual Tubers)?}
    \begin{itemize}
        \item Yes
        \item No
    \end{itemize}

\item \textbf{1.2 Do you watch the AI-driven VTuber, Neuro-sama?}
    \begin{itemize}
        \item Yes
        \item No
    \end{itemize}

\item \textbf{1.3 Approximately how often do you watch Neuro-sama's streams or related video content?}
    \begin{itemize}
        \item Almost every day
        \item 3-5 times a week
        \item 1-2 times a week
        \item A few times a month
        \item Occasionally
    \end{itemize}

\item \textbf{1.4 Before watching Neuro-sama, were you already a regular viewer of human-piloted VTubers?}
    \begin{itemize}
        \item Yes
        \item No
    \end{itemize}

\item \textbf{1.5 How did you first learn about Neuro-sama? (Select all that apply)}
    \begin{itemize}
        \item[$\square$] Algorithm recommendation on a video platform (e.g., YouTube/Twitch)
        \item[$\square$] Short clips on social media (e.g., Twitter, Tik Tok)
        \item[$\square$] Recommendation from a friend or an online community member
        \item[$\square$] Discussions on a forum like Reddit
        \item[$\square$] News or tech media reports
        \item[$\square$] Other
    \end{itemize}

\item \textbf{1.6 Please rate the importance of the following factors in initially attracting you to watch Neuro-sama. (Scale: 1 - Not at all important, 5 - Extremely important)}
    \begin{itemize}
        \item Curiosity about AI/LLM technology
        \item Funny/chaotic clips circulating online
        \item Her unique virtual avatar design
        \item The unpredictable and surprising atmosphere of the streams
        \item The interesting way the community interacts with the AI
        \item The novelty of a streamer having no ``human behind the curtain''
    \end{itemize}

\item \textbf{1.7 Was there a specific moment or interaction that made you decide to become a regular viewer instead of just a casual one? If so, please describe it.}

\item \textbf{1.8 (Optional) Compared to the human VTubers you watched previously, how do you perceive Neuro-sama?} \\
(Scale: Much lower than human VTubers, Lower than human VTubers, About the same, Higher than human VTubers, Much higher than human VTubers)
    \begin{itemize}
        \item Entertainment value of the content
        \item Level of interactivity with the audience
        \item The unpredictability / surprise factor of the content
    \end{itemize}

\item \textbf{2.1 Please rate your agreement with the following statements. (Scale: 1 - Strongly disagree, 5 - Strongly agree)}
    \begin{itemize}
        \item I pay close attention to Neuro-sama's behaviors and response patterns.
        \item I feel like I have a good understanding of Neuro-sama's ``personality,'' and can sometimes predict how ``she'' will react to a certain question.
        \item I often think about why Neuro-sama says certain things, or what the underlying logic of ``her'' system might be.
        \item Neuro-sama's words and actions make me feel that ``it'' has a coherent and unique ``personality.''
    \end{itemize}

\item \textbf{2.2 Please rate your agreement with the following statements. (Scale: 1 - Strongly disagree, 5 - Strongly agree)}
    \begin{itemize}
        \item I like Neuro-sama as a ``streamer.''
        \item Watching Neuro-sama's streams makes me feel relaxed and comfortable.
        \item When Neuro-sama is praised for performing well, I feel happy too.
        \item If Neuro-sama runs into trouble due to a technical issue or being ``tricked'' by viewers, I feel a little ``sad'' or sympathetic.
    \end{itemize}

\item \textbf{2.3 Please rate your agreement with the following statements. (Scale: 1 - Strongly disagree, 5 - Strongly agree)}
    \begin{itemize}
        \item During a stream, I often feel the urge to ask Neuro-sama a question or express my opinion via chat or a Super Chat.
        \item When I come across an interesting topic, I wonder what Neuro-sama would have to say about it.
        \item I would recommend or share streams or clips of Neuro-sama with my friends.
        \item If Neuro-sama holds a special online event (like a Dev Stream or a birthday stream), I would be interested in participating.
    \end{itemize}

\item \textbf{2.4 When you think about your ``relationship'' with Neuro-sama, which of the following descriptions feel most accurate? (Select all that apply)}
    \begin{itemize}
        \item[$\square$] A chaotic performer who brings joy
        \item[$\square$] An AI that I am helping to ``train'' or ``raise''
        \item[$\square$] A digital daughter that needs to be cared for and interacted with
        \item[$\square$] A program with unknown limits to be tested and explored
        \item[$\square$] A virtual friend who provides companionship
        \item[$\square$] A tech project created by a talented developer
        \item[$\square$] Other
    \end{itemize}

\item \textbf{2.5 In the Neuro-sama fan community, what is your favorite or most memorable ``meme''? Why do you think this meme is important to the community?}

\item \textbf{3.1 Have you ever paid for content related to Neuro-sama (e.g., sending a Super Chat or buying a channel membership)?}
    \begin{itemize}
        \item Yes
        \item No
    \end{itemize}

\item \textbf{3.2 (Optional) What is the approximate frequency of your payments to Neuro-sama?}
    \begin{itemize}
        \item I pay during almost every stream
        \item A few times a month
        \item Only for special occasions (e.g., birthday stream)
        \item I have paid before, but no longer do so
        \item Other
    \end{itemize}

\item \textbf{3.3 (Optional) What are your primary motivations for paying? (Please select the most important 1-2 options)}
    \begin{itemize}
        \item[$\square$] To test the AI's response to a specific question or command
        \item[$\square$] To have my comment highlighted and influence the stream's content
        \item[$\square$] To reward a particularly funny or impressive moment that just happened
        \item[$\square$] To express my affection and support for the ``character'' of Neuro-sama
        \item[$\square$] To express my support for the developer (Vedal) and his work
        \item[$\square$] As a way of participating in a community celebration or ritual
        \item[$\square$] Other
    \end{itemize}

\item \textbf{3.4 (Optional) Please briefly describe a specific situation when you recently paid for Neuro-sama. What was happening in the stream?}

\item \textbf{4.1 What is your age?}
    \begin{itemize}
        \item Under 18
        \item 18-24
        \item 25-34
        \item 35-44
        \item 45+
        \item Prefer not to say
    \end{itemize}

\item \textbf{4.2 What is your gender identity?}
    \begin{itemize}
        \item Male
        \item Female
        \item Non-binary
        \item Other
        \item Prefer not to say
    \end{itemize}

\item \textbf{4.3 What is your geographical region?}
    \begin{itemize}
        \item North America
        \item Europe
        \item East Asia
        \item Southeast Asia
        \item Latin America
        \item Other
        \item Prefer not to say
    \end{itemize}

\item \textbf{(Optional) Is there anything else you would like to add on this topic?}
\end{enumerate}

\section{Codebook of Interview}
\label{Appendix: codebook}
\begin{enumerate}[label=(\arabic*), font=\bfseries, itemsep=5pt, topsep=5pt]
  \item Sources of unexpected moments
  \begin{enumerate}[label=(\alph*), font=\normalfont, itemsep=2pt]
    \item Breaking Through Technical Limitations
    \item Entertaining interactions with the community
    \item Dialogues Showing Emotional \& Philosophical Depth
    \item A Sense of the Future Technology
    \item Novel Concept of an "AI VTuber"
    \item No Specific Moments of Surprise
  \end{enumerate}

  \item Perceived differences in unexpected moments: AI vs. Human 
  \begin{enumerate}[label=(\alph*), font=\normalfont, itemsep=2pt]
    \item Stems from technical characteristics
    \item Stems from community shaping
    \item No Perceived Difference
  \end{enumerate}

  \item Importance of Neuro-sama having no Nakanohito
  \begin{enumerate}[label=(\alph*), font=\normalfont, itemsep=2pt]
    \item Important, because it won't suddenly "collapse" or disappear
    \item Important, because it prevents doxxing of the Nakanohito
    \item Important, because it ensures the purity of the Role-Playing experience
    \item Not important
  \end{enumerate}

  \item Views on Neuro-sama's emotions
  \begin{enumerate}[label=(\alph*), font=\normalfont, itemsep=2pt]
    \item Has emotions like a real person  
    \item Has no emotions rationally, but is experienced as if having emotions emotionally  
    \item Uncertain, but tendency toward having emotions  
    \item Does not have emotions  
  \end{enumerate}
  
  \item Interviewees' emotions toward Neuro-sama
  \begin{enumerate}[label=(\alph*), font=\normalfont, itemsep=2pt]
    \item Emotional, similar to fictional characters
    \item Emotional, similar to real people
    \item Emotional, stronger than for real people
    \item No Emotional Investment
  \end{enumerate}
  
  \item Views on Neuro-sama's personality
  \begin{enumerate}[label=(\alph*), font=\normalfont, itemsep=2pt]
    \item The personality is understood as inherent to Neuro-sama and continuously evolving.
    \item The personality is viewed not as an inherent attribute of Neuro-sama, but as a construct collectively formed by the community and the AI itself.
    \item The personality is the result of LLM responding to prompts.
    \item The participant is hesitant to call it a true personality, viewing it as too simple and lacking human complexity.
  \end{enumerate}
  
  \item Description of the relationship with Neuro-sama
  \begin{enumerate}[label=(\alph*), font=\normalfont, itemsep=2pt]
    \item A child
    \item A companion
    \item A source of emotional/spiritual support
    \item A performer
    \item A technology project
  \end{enumerate}
\end{enumerate}

\section{Interview Recruitment Material}
Dear members of The Swarm!

We are an academic team dedicated to HCI research. We sincerely invite devoted viewers of Neuro-sama to participate in a supplementary semi-structured academic interview about AI VTuber experiences.

For this interview, we particularly hope to invite viewers who have both of the following viewing experiences:
1. Familiarity with and frequent viewing of Neuro-sama's livestreams or clip content
2. Current or past following of human VTubers (excluding Vedal)

The interview format will be a one-on-one online voice or text conversation, lasting approximately 20-30 minutes.

To thank you for your valuable time and sharing, after completing the interview, we will provide a subscription to Neuro-sama (@vedal987) on Twitch or an equivalent cash amount as a token of appreciation.

Like the questionnaire, this interview will be completely anonymous. All collected data will be strictly confidential and used only for academic research and publication.

If you meet the requirements and are interested in this interview, please contact me through [registration link] to sign up. We plan to recruit 8-12 interviewees and very much look forward to having an in-depth conversation with you!

\section{Interview Informed Consent Statement}
Before we begin the interview, I want to ensure you understand the nature of this research study:

This research is conducted by our academic team, aiming to understand how audiences interact with AI-driven VTubers like Neuro-sama and explore the unique relationships and community cultures that emerge.

Your participation is completely voluntary. You may choose not to answer any question or leave the study at any time without penalty.

The interview will last approximately 20-30 minutes and will be audio-recorded with your permission. All your information will be kept confidential - your identity will be protected with pseudonyms in any publications, and recordings will be securely stored and eventually destroyed.

The audio recording will be stored in an encrypted, password-protected file accessible only to the core research team. During transcription, all personally identifying information will be removed and replaced with pseudonyms. The original audio recordings will be permanently destroyed after the transcripts have been verified for accuracy, or within five years of the study's completion, whichever comes first.

There are minimal risks to participating, primarily related to potential confidentiality breaches, which we take extensive measures to prevent.

If you have any questions about your rights as a research participant or concerns about the study, you can contact us at [contact information].

By continuing with this interview, you acknowledge that you have understood this information and consent to participate in this research study and to be audio-recorded. Do you have any questions before we proceed?

\section{Prompt Template}
\label{appendix:prompt_template}
\begin{figure*}[h]
\centering
\begin{bluebox}[
  Prompt Template: Chat Messages Coding
]
[System]
\newline You are an expert research assistant in Social Computing. Your mission is to analyze the provided live stream chat message and classify its primary interactive intent into ONE of the following categories.
\newline The following chat message comes from the community of a VTuber \{vtuber\_name\}.
\{vtuber\_background\}.
\newline
\newline [Coding Scheme]
\newline A-POS (Positive Affect): Expresses positive emotions, support, affection, or encouragement towards the streamer or others.
\newline A-NEG (Negative Affect): Expresses genuinely harmful negative emotions, such as malicious teasing, insulting, belittling, or aggressive verbal attacks.
\newline Q-CMD (Question/Command): Asks a meaningful question seeking information or issues a substantive command/request to the streamer. The question or command should have clear intent beyond simple reactions.
\newline R-GEN (General Reaction): Simple expressions, reactions, chat rituals, or statements that show viewer engagement (e.g., "lol", "I'll come", "pog", "LMAO", "let's go", emotes, memes, copypasta, or common chat patterns).
\newline C-SOC (Social/Community): Interacts directly with other chat users, typically using "@" or responding to others' messages.
\newline N/A (Not Applicable): The message is gibberish, non-English, spam, or its intent is impossible to determine.
\newline
\newline [Message to Classify]
\newline \{chat\_message\}
\newline
\newline [Instruction]
\newline Based on the coding scheme, what is the single best code for the message above? Respond with ONLY the code ID (e.g., A-POS, Q-CMD, N/A). Do not provide any explanation or extra text.

\end{bluebox}
\caption{prompt template for coding Chat messages. The template is dynamically populated for each Chat instance being analyzed, where \texttt{\{vtuber\_name\}} is replaced with the streamer's name, \texttt{\{vtuber\_background\}} provides context about the streamer, \texttt{\{chat\_message\}} contains the full text of the Chat.}
\label{prompt: coding_chat} 
\end{figure*}

\begin{figure*}[h]
\centering
\begin{bluebox}[
  Prompt Template: SuperChat Coding
]
[System]
\newline Analysis Task: VTuber Live Stream Interaction
\newline
\newline [Background]
\newline You are analyzing a 90-second video clip from a live stream featuring \{vtuber\_name\}, an VTuber. \{vtuber\_background\}. Please note that the Superchat prompt used by \{vtuber\_name\} is a voice prompt, so the user's Superchat will be read out directly. You need to pay attention to this. If you hear the same voice as Superchat, it means that this is the user's Superchat.
\newline
\newline [Context of this Clip]
\newline Super Chat Message: "\{message\}"
\newline Time Sent: The Super Chat was sent at approximately the 30-second mark of this video clip (originally at stream timestamp \{timestamp\}).
\newline
\newline [Your Goal \& Questions to Answer]
\newline Your mission is to analyze the interaction in the video clip and provide a structured response. Focus on the relationship between the streamer's actions and the viewer's SuperChat. Was this Super Chat PROACTIVE or REACTIVE?
\newline - PROACTIVE: The user is initiating a new question, a new topic, or making a request that is not directly related to what the streamer was just talking about.
\newline - REACTIVE: The user is responding directly to something the streamer just said or did in the moments leading up to the Super Chat.
\newline - BOTH: It contains elements of both. For example, reacting to a topic but using it to ask a new, expansive question.
\newline - UNCLEAR: It is not possible to determine the initiative from the clip.
\newline
\newline [Output Format Specification]
\newline You MUST provide your analysis ONLY within the following structured block. Do not include any other text, greetings, or explanations outside of this block.
\newline [ANALYSIS\_START]
\newline INITIATIVE: <Fill in PROACTIVE, REACTIVE, BOTH, or UNKNOWN>
\newline [ANALYSIS\_END]
\end{bluebox}
\caption{The prompt template used to instruct the LLM for the contextual coding of SuperChat messages. The template is dynamically populated for each SuperChat instance being analyzed, where \texttt{\{vtuber\_name\}} is replaced with the streamer's name, \texttt{\{vtuber\_background\}} provides context about the streamer, \texttt{\{message\}} contains the full text of the SuperChat, and \texttt{\{timestamp\}} indicates the original time the message was sent during the livestream.}
\label{prompt: coding_superchat}
\end{figure*}

\end{document}